\documentclass[preprint,times]{elsarticle}
\pdfoutput=1

\usepackage{amsmath}  
\usepackage{amssymb}
\usepackage{subfigure}
\usepackage{ecrc}
\usepackage{rotating}
\usepackage{xcolor}
\usepackage{listings}
\usepackage{latexsym}
\usepackage{bm}
\usepackage{bbm}
\usepackage{floatrow}
\usepackage{breqn}
\newfloatcommand{capbtabbox}{table}[][\FBwidth]
\usepackage{tikz}
\definecolor{mygreen}{RGB}{28,172,0} 
\definecolor{mylilas}{RGB}{170,55,241}

\DeclareMathOperator*{\inf2}{inf}
\DeclareMathOperator*{\sup2}{sup}
\volume{00}

\firstpage{1}
\runauth{}

\begin{document}
\lstset{language=Matlab,%
    breaklines=true,%
    morekeywords={matlab2tikz},
    keywordstyle=\color{blue},%
    morekeywords=[2]{1}, keywordstyle=[2]{\color{black}},
    identifierstyle=\color{black},%
    stringstyle=\color{mylilas},
    commentstyle=\color{mygreen},%
    showstringspaces=false,
    numbers=left,%
    numberstyle={\tiny \color{black}},
    numbersep=9pt, 
    emph=[1]{for,end,break},emphstyle=[1]\color{red}, 
}
\begin{frontmatter}

\title{A coalescence criterion for porous single crystals}

\author[CEA]{J.~Hure\corref{cor1}}

\cortext[cor1]{Corresponding author}
\address[CEA]{DEN-Service d'Etudes des Mat\'eriaux Irradi\'es, CEA, Universit\'e Paris-Saclay, F-91191, Gif-sur-Yvette, France}

\begin{abstract}
  Voids can be observed at various scales in ductile materials, frequently of sizes lower than the grain size, leading to porous single crystals materials. Two local deformation modes of porous ductile (single crystals) materials have been identified and referred to as void growth and void coalescence, the latter being characterized by strong interactions between neighboring voids. A simple semi-analytical coalescence criterion for porous single crystals with periodic arrangement of voids is proposed using effective isotropic yield stresses associated with a criterion derived for isotropic materials. An extension of the coalescence criterion is also proposed to account for shear with respect to the coalescence plane. Effective yield stresses are defined using Taylor theory of single crystal deformation, and rely ultimately on the computation of average Taylor factors. Arbitrary sets of slip systems can be considered. The coalescence criterion is assessed through comparisons to numerical limit-analysis results performed using a Fast-Fourier-Transform based solver. A good agreement is observed between the proposed criterion and numerical results for various configurations including different sets of slip systems (Face-Centered-Cubic, Hexagonal-Close-Packed), crystal orientations, void shapes and loading conditions. The competition between void growth and void coalescence is described for specific conditions, emphasizing the strong influence of both crystal orientation and void lattice, as well as their interactions.
\end{abstract}


\end{frontmatter}

\section{Introduction}

Ductile fracture through void growth to coalescence is one of the failure mode of metal alloys. Early observations of the effect of porosity on fracture \cite{tipper,puttick} have motivated researches on voids evolution under mechanical loading \cite{mcclintock,ricetracey}, emphasizing the major roles of hydrostatic stress and porosity on the mechanical behavior of porous materials. Homogenized models for isotropic porous materials have subsequently been proposed by Gurson \cite{gurson} and extended by Tvergaard and Needleman \cite{tvergaardneedleman} into the so-called GTN model which is still widely used. Seminal porous unit-cells computations \cite{koplik} have put forward two deformation modes for porous materials with periodic arrangement of voids, known as void growth (implicitly considered in the Gurson model), and void coalescence (corresponding to localized plastic flow). A coalescence criterion was then proposed by Thomason \cite{thomason85a} based on a similar approach as the one followed by Gurson. Several homogenized models have since then been proposed to incorporate the effects of void shape, anisotropy, strain-hardening, and used to simulate ductile tearing. Detailed reviews about the underlying physical mechanisms of ductile fracture through void growth to coalescence, homogenized models and their numerical implementation can be found in \cite{besson2010,benzergaleblond,BLNT,pineaureview}.

Based on unit-cells simulations (see, \textit{e.g.}, \cite{daehli,legarth} for recent studies and references therein), plastic anisotropy has been incorporated into homogenized models for porous materials, mainly through Hill's orthotropic yield criterion, for growth \cite{benzergaanisotrope,monchiet,benzergagld,morinellipse,danas} and more recently for coalescence \cite{morinthese,keralavarma,gallican}. However, void growth and coalescence have been much less studied in single crystals, where strong plastic anisotropy arises from well-defined sets of slip systems, while associated homogenized models will be ultimately required to model macroscopic mechanical behavior up to ductile tearing. Typical examples of potential applications are single crystals used as structural components such as Nickel-based superalloys in aerospace applications \cite{dennisthese}, or polycrystalline aggregates with voids lower than the grain size arising from submicrometric particles \cite{daly}, vacancies condensation prior \cite{cawthorne} or under deformation \cite{rosi}. Void growth in single crystals has been described experimentally, \textit{e.g.} by Cr\'epin \textit{et al.} \cite{crepin} and Gan \textit{et al.} \cite{gan2}, emphasizing the strong interactions between crystallographic orientations and void growth shapes as well as lattice rotations around voids. Crystal-plasticity constitutive equations \cite{asaro} have been used in finite-element simulations of porous unit-cells \cite{koplik} to assess voids evolution in single crystals under mechanical loading. O'Regan \textit{et al.} \cite{oregan}, Orsini \& Zikry \cite{orsini} and Potirniche \textit{et al.} \cite{potirniche} performed 2D simulations which showed a strong influence of crystal orientation and void arrangement on void growth and coalescence, especially for low stress triaxialities. Additional crystal-plasticity three-dimensional simulations \cite{liu,ha,yerra} confirmed the first observations and detailed the influence of crystal orientation on void coalescence. Previous studies used (almost)-rate-independent crystal plasticity, but rate-dependent constitutive equations have also been used to assess creep behavior of porous single crystals \cite{srivastava1,srivastava2}, where either void coalescence or void collapse can be observed. In addition, Discrete-Dislocations-Dynamics (DDD) simulations \cite{segurado2009,segurado2010,chang2015} provided valuable details about anisotropic dislocations motion around voids for low values of applied strain. Some mechanistic studies about void growth in single crystals have also addressed size effects in addition to crystallographic orientation effects, mainly through Molecular Dynamics (\textit{e.g.} \cite{traiviratana}) or strain-gradient crystal plasticity finite-element simulations (\textit{e.g.} \cite{shu,borg2008}), which are outside the scope of this paper.

\begin{figure}[H]
\centering
\subfigure[]{\includegraphics[height = 5cm]{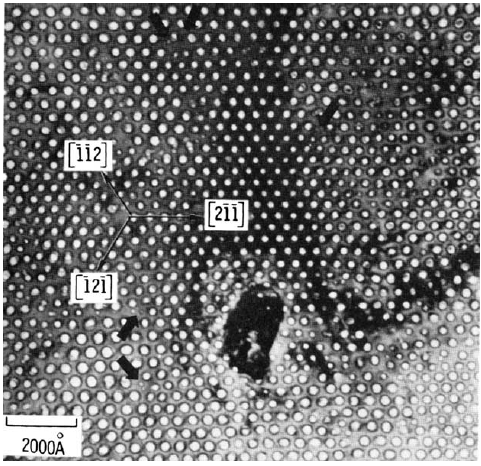}}
\hspace{1cm}
\subfigure[]{\includegraphics[height = 5cm]{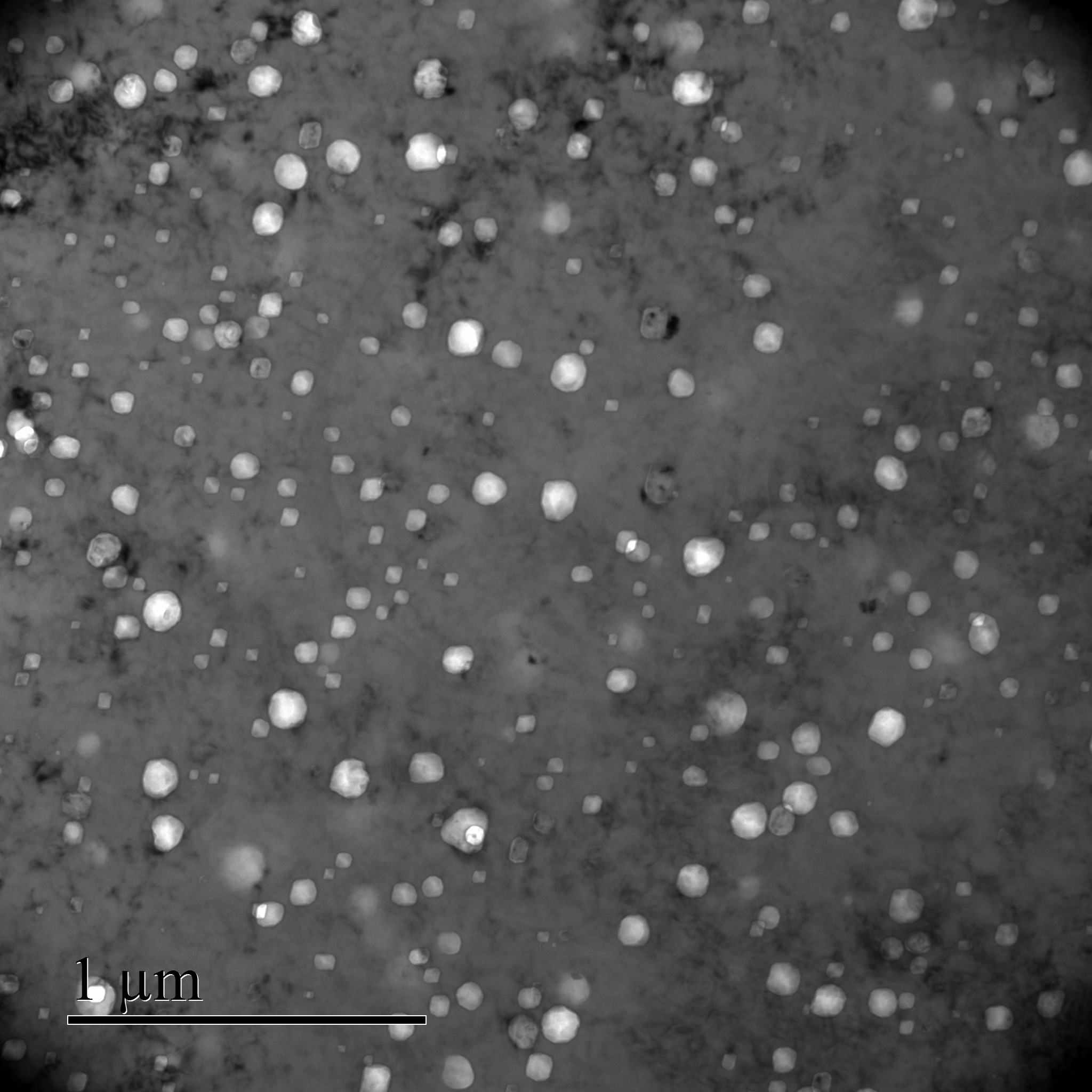}}
\caption{Transmission Electron Microscope (TEM) images showing (a) regular nanovoids lattice in Niobium irradiated with 7.5 MeV Ta$^+$ ions at 800$^{\circ}$C \cite{ghoniem} and (b) random distribution of nanovoids in an austenitic stainless steel irradiated with 2MeV Fe$^{3+}$ at 600$^{\circ}$C (Courtesy of P.O. Barrioz).}
\label{fig0}
\end{figure}

Three-dimensional homogenized models for porous single crystals were proposed only recently. Han \textit{et al.} \cite{xuhan} developed a yield criterion for porous single crystals assuming spherical voids and rate-independent crystal plasticity, leading to a crystal plasticity model where the critical resolved shear stress depends on porosity and stress triaxiality. Following similar hypotheses, another yield criterion was proposed in \cite{paux}, furthermore validated through comparisons to unit-cells simulations and extended to account for hardening \cite{paux2}. Based on the variational approach, a more general yield criterion for porous single crystals was derived in \cite{mbiakop2} considering distributions of ellipsoidal voids and rate-dependent crystal plasticity. For spherical voids, the three yield criteria \cite{xuhan,paux,mbiakop2} lead to rather similar results, for rate-independent FCC materials. For the sake of completeness, it should be noticed that a homogenized model for rate-dependent porous single crystals was proposed by \cite{srivastava3}, while dynamic effects have been studied in \cite{nguyen}. Based on the criterion proposed in \cite{xuhan}, a complete set of constitutive equations has been proposed by Ling \textit{et al.} \cite{chaoling} to model porous crystals at finite strains, including void growth and hardening, and validated against unit-cells simulations of initially spherical voids of simple cubic lattice in a FCC material. Song \& Ponte-Casta{\~ n}eda \cite{song1,song2} proposed a set of constitutive equations for viscoplastic porous single crystals based on the variational approach, validated against comparisons to unit-cells simulations performed in \cite{srivastava2}.

Porous unit-cells simulations with crystal plasticity have shown that a coalescence deformation mode occurs at large porosities \cite{yerra,chaoling}, associated with localized plastic flow and elastic unloading. Interestingly, coalescence has been observed to occur easily where the intervoid ligament is parallel to the maximal growth direction \cite{liu}, demonstrating the strong coupling between crystallographic orientation and void arrangement at coalescence. None of the homogenized models for porous single crystals are able to describe coalescence, as clearly shown by the comparisons to unit-cells simulations in \cite{chaoling,song2}. Significant progresses on the theoretical derivation and numerical validation of coalescence criteria for porous materials (with periodic voids arrangement) have been made recently, based on the seminal work of Thomason \cite{thomason85a} and phenomenological extensions based on it \cite{benzerga02,pardoen,tekoglushear}. Rigorous coalescence criteria have been proposed for isotropic matrix materials \cite{thomasonnew1,thomasonnew2,hurebarrioz}, incorporating the presence of shear loading with respect to the coalescence plane \cite{torki,torki2017}, and extended to account for plastic anisotropy through Hill's criterion \cite{morinthese,keralavarma,gallican}. Yerra \textit{et al.} \cite{yerra} proposed heuristic extensions to Thomason original criterion - derived for isotropic materials - to describe coalescence of porous single crystals, but to date no coalescence criterion has been proposed dedicated to porous single crystals.

The aim of this study is thus to propose a coalescence criterion for porous single crystals, accounting for slip systems and crystallographic orientation of the matrix material surrounding voids, void shapes and lattices, restricting to periodic arrangement of voids. A detailed statement of the problem as well as the main assumptions are described in Section~2. In Section~3, a coalescence criterion is proposed based on limit-analysis, homogenization and crystal plasticity. The proposed criterion is validated against comparisons to numerical limit-analysis simulations in Section~4. The competition between void growth and void coalescence is described for specific conditions in Section~5, emphasizing the interplay between both crystal orientation and void lattice.

\section{Problem statement}

\begin{figure}[H]
\centering
\subfigure[]{\includegraphics[height = 4cm]{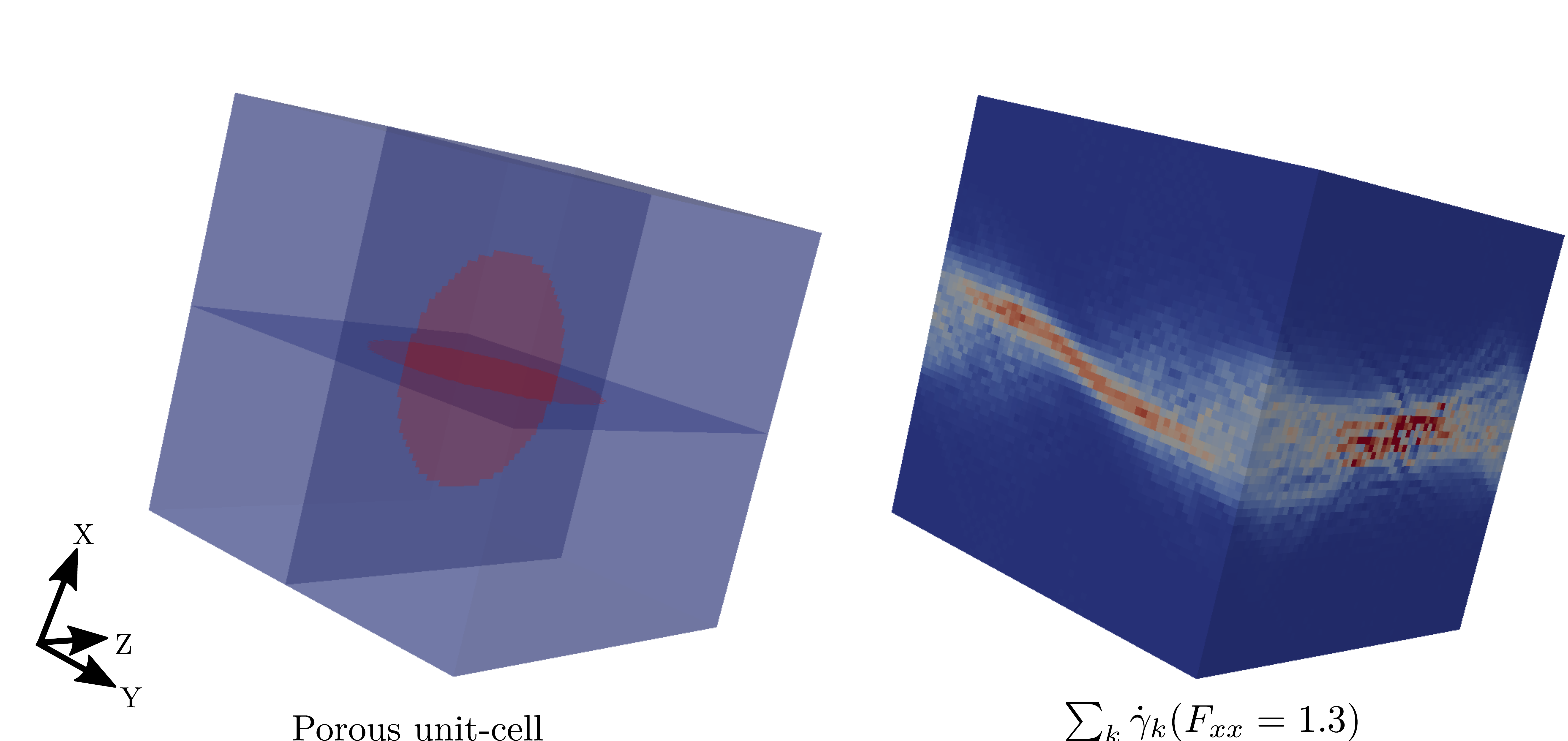}}
\subfigure[]{\includegraphics[height = 4cm]{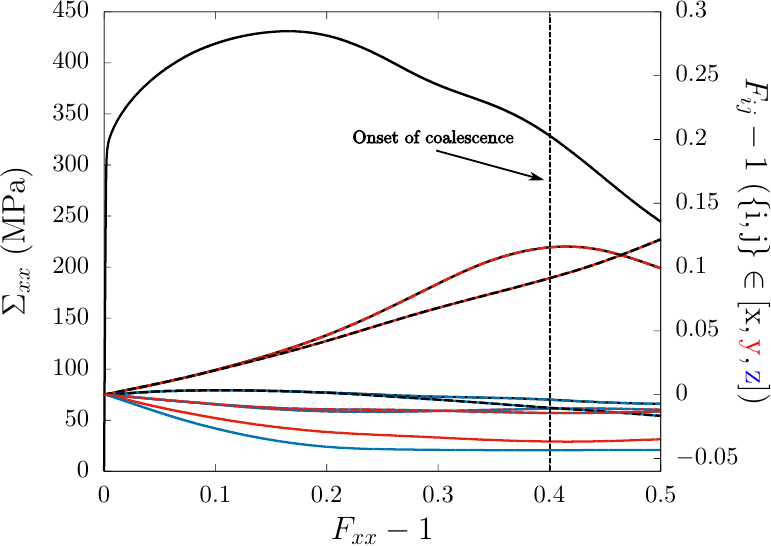}}
\caption{Finite-strain unit-cell simulation (performed with \texttt{AMITEX\_FFTP} FFT-based solver \cite{amitex}) of a porous FCC single crystal (with crystallographic orientation  $x=[\bar{1}25]$  $y=[1\bar{2}1]$ $z=[210]$) under axisymmetric loading conditions (with stress triaxiality $T=2$) with periodic boundary conditions. Constitutive equations can be found in \cite{chaoling}.  $\mathrm{(a)}_1$ \textbf{Simple cubic array of voids} (of porosity $f=0.1$) (b) Evolution of macroscopic (through volume averaging) axial stress $\Sigma_{xx}$ and deformation gradients $F_{ij}$ as a function of axial deformation gradient $F_{xx}$. $\mathrm{(a)}_2$ Snapshot of the cumulative plastic slips in the coalescence regime (arbitrary units).}
\label{fig1}

\end{figure}

\begin{figure}[H]
\centering
\subfigure[]{\includegraphics[height = 4cm]{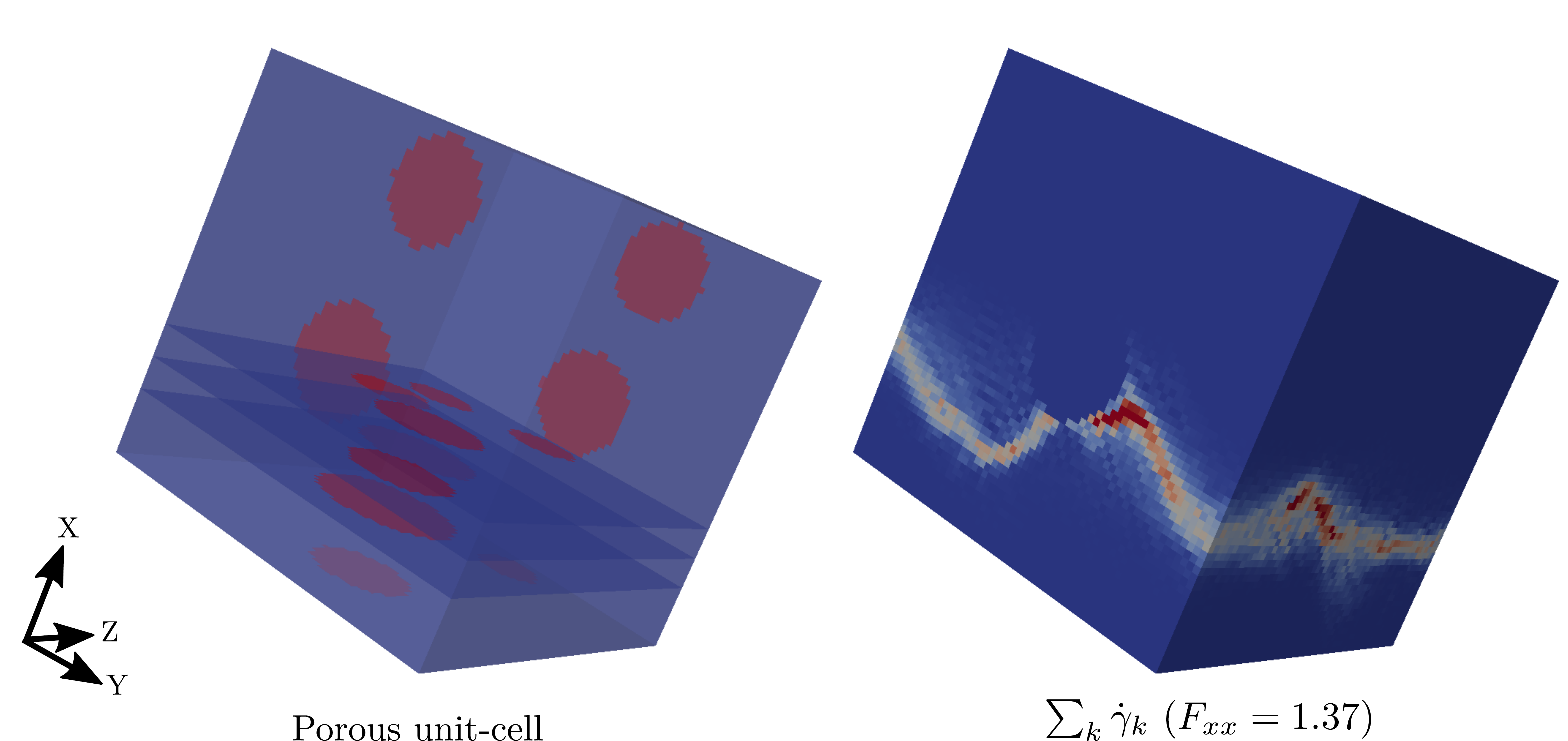}}
\subfigure[]{\includegraphics[height = 4cm]{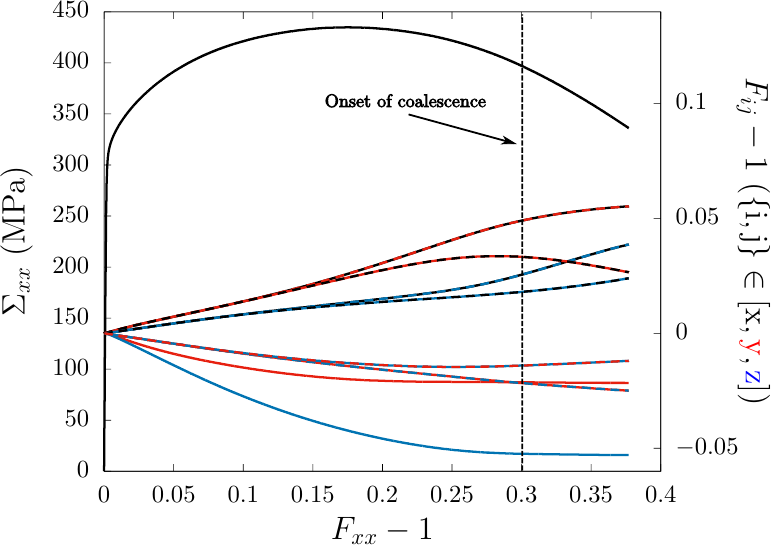}}
\caption{Finite-strain unit-cell simulation (performed with \texttt{AMITEX\_FFTP} FFT-based solver \cite{amitex}) of a porous FCC single crystal (with crystallographic orientation  $x=[\bar{1}25]$  $y=[1\bar{2}1]$ $z=[210]$) under axisymmetric loading conditions (with stress triaxiality $T=2$) with periodic boundary conditions. Constitutive equations can be found in \cite{chaoling}.  $\mathrm{(a)}_1$ \textbf{(Pseudo-)random distributions of (5) voids} (of porosity $f=0.1$) (b) Evolution of macroscopic (through volume averaging) axial stress $\Sigma_{xx}$ and deformation gradients $F_{ij}$ as a function of axial deformation gradient $F_{xx}$. $\mathrm{(a)}_2$ Snapshot of the cumulative plastic slips in the coalescence regime (arbitrary units).}
\label{fig2}
\end{figure}

Typical results of finite-strain single crystals porous unit-cells simulations under periodic boundary conditions are recalled on Figs.~\ref{fig1} and \ref{fig2} (see captions for details), for both simple cubic and (pseudo-)random arrangement of voids, that correspond to the experimental situations shown on Fig.~\ref{fig0}. In both cases, the evolution of the axial stress shows first hardening (due to the hardening of the matrix material around the voids) and then softening due to the increase of porosity. In the softening regime, a coalescence mode defined as localized plastic flow in a layer of typical width sets by void size occurs for both periodic and non-periodic void microstructure, as shown by the vanishing deformation gradient rates in transverse directions. Moreover, shear stresses with respect to the coalescence plane may play an important role, as out-of-plane deformation gradients do not vanish at coalescence. Finally, these simulations indicate that the coalescence plane may depend on both void microstructure and crystallographic orientation, as clearly shown in Fig.~\ref{fig2} for (pseudo-)random arrangement of voids.

The comparison of results presented in Figs.~\ref{fig1} and \ref{fig2} shows that, although a coalescence deformation regime occurs in both cases, voids distribution (random \textit{vs.} periodic) may have a non-negligible effect on both coalescence stress and localization of plastic flow. This study is restricted to developing and validating a coalescence criterion for periodic arrangement of voids in single crystals. This corresponds on one hand to some experimental observations (Fig.~\ref{fig0}a), and on the other hand to a simplifying assumption allowing to derive analytical coalescence criteria \cite{thomasonnew1} but requiring assessment for the coalescence criterion to be used for random distributions of voids.
 
\section{Derivation of the coalescence criterion}

Underline $\underline{A}$, bold $\textbf{A}$ and double-struck $\mathbbm{A}$ symbols refer to vectors, second-order and fourth-order tensors, respectively. A cartesian orthonormal basis $\{\underline{e}_1,\underline{e}_2,\underline{e}_3\}$ and a cylindrical orthonormal basis $\{\underline{e}_r,\underline{e}_{\theta},\underline{e}_z=\underline{e}_3\}$ along with coordinates $\{x,y,z \}$ and $\{r,\theta,z \}$, respectively, are used. Volume averaging is denoted as $<.>_{\Omega} = [\int_{\Omega} .\ d\Omega]/vol{\Omega}$.

\subsection{Theoretical background}

\subsubsection{Crystal plasticity}

A rate-independent small strain crystal plasticity framework is used to model single crystals for the derivation of the coalescence criterion, without elasticity\footnote{The framework of limit-analysis used in the following to estimate the coalescence criterion states that limit-loads do not depend on elasticity, thus rigid-plastic material can be considered.} nor hardening \cite{bishop}. Plastic flow is assumed to be only the consequence of dislocation glide in well-defined slip systems. A slip system $k$ is fully described with the gliding direction $\underline{m}^k$ and plane of normal $\underline{n}^k$. Symmetric Schmid tensor reads:
\begin{equation}
  \bm{\mu}^k = \frac{1}{2}\left(\underline{m}^k \otimes \underline{n}^k + \underline{n}^k \otimes \underline{m}^k \right)
\end{equation}
Shear stress acting on each slip system can be computed using Cauchy stress tensor $\bm{\sigma}$ and Schmid tensor as:
\begin{equation}
  \tau^k = \bm{\sigma}:\bm{\mu}^k
  \end{equation}
where $\tau^k$ is the resolved shear stress in the slip system $k$. Schmid criterion is used for glide to occur in a given slip system:
\begin{equation}
   \tau^k - \tau_0  \leq 0 \ \ \ \ \  \dot{\gamma}_k \geq 0 \ \ \ \ \ (\tau^k - \tau_0)\dot{\gamma}_k = 0
\end{equation}
where $\tau_0$ is the critical resolved shear stress (assumed to be the same for all slip systems) and $\gamma_k$ the plastic slip in the system $k$. Finally, the (plastic) strain rate tensor can be written as:
\begin{equation}
  \bm{\dot{\epsilon}} = \sum_k \dot{\gamma}_k \,\bm{\mu}^k
\end{equation}
Regularized versions are usually used for numerical purposes, either considering a viscoplastic version of Schmid's law \cite{hutchinson76} or a single yield criterion \cite{arminjon,gambin}, to remove the indeterminacy in the case of multi-slip configurations for an applied strain rate tensor. For unregularized crystal plasticity, Taylor's minimum shear principle \cite{taylor38} or equivalent Bishop-Hill maximal plastic work principle \cite{bishop} can be used to select the active slip systems. Taylor factor is usually defined to quantify the total amount of plastic slips:
\begin{equation}
  M = \frac{\sum_k \dot{\gamma}_k}{d_{eq}}
  \label{taylorfactor}
\end{equation}
where $d_{eq}=\sqrt{(2/3)\dot{\bm{\epsilon}}:\dot{\bm{\epsilon}}}$ is the von Mises equivalent (plastic) strain rate. Taylor factor is  widely used to estimate the yield stress of polycrystalline aggregates by averaging over the different crystallographic orientations.
\subsubsection{Homogenization and Limit-Analysis}

Most of the coalescence criteria for porous materials proposed in previous studies were derived through limit-analysis framework along with homogenization. A detailed description of limit-analysis can be found in \cite{benzergaleblond}. Key ingredients that will be used in the following are described hereafter. A unit-cell $\Omega$ under homogeneous strain-rate or periodic boundary conditions is considered, with an incompressible plastic matrix material. Macroscopic stress $\bm{\Sigma}$ and strain rate \textbf{D} tensors are related to their microscopic counterparts by volume averaging:
\begin{equation}
\bm{\Sigma} = <\bm{\sigma} >_{\Omega} \ \ \ \ \ \ \ \ \ \ \textbf{D} = < \textbf{d} >_{\Omega}
\label{macroscopic}
\end{equation}
with $\bm{\sigma}$ the Cauchy stress and $\textbf{d}$ the microscopic strain rate tensor. Hill-Mandel lemma reads:
\begin{equation}
< \bm{\sigma}:\textbf{d} >_{\Omega} =  \bm{\Sigma}:\textbf{D}
\end{equation}
which allows making the transition from microscale to the macroscale. The so-called upper-bound theorem of limit-analysis, based on the principle of maximal plastic dissipation, is \cite{benzergaleblond}:
\begin{equation}
  \textcolor{black}{ \bm{\Sigma}:\textbf{D} = \Pi(\textbf{D})  = \inf2_{\underline{v} \in K(\textbf{D})} < \sup2_{\sigma^{\star} \in C} \bm{\sigma}^{\star}:\textbf{d} >_{\Omega}          }
  \label{eqanalyselimite}
\end{equation}
where $K(\textbf{D})$ is the subset of velocity field kinematically admissible, \textit{i.e.}, compatible with boundary conditions $<\textbf{d}(\underline{v})>_{\Omega} = \textbf{D}$ and verifying the property of incompressiblity  $\mathrm{tr}(\textbf{d})=0$, and $C$ the microscopic plastic strength domain. $\Pi(\textbf{D})$ will be referred to as the macroscopic plastic dissipation. The macroscopic plastic strength domain is obtained from Eq.~\ref{eqanalyselimite} by the equation:
\begin{equation}
\bm{\Sigma} = \frac{\partial \Pi(\textbf{D}) }{\partial \textbf{D}}
\label{eqanalyselimite2}
\end{equation}
Evaluating the macroscopic plastic strength domain - referred to as the coalescence criterion when boundary conditions compatible with coalescence are used - through Eqs.~\ref{eqanalyselimite} and \ref{eqanalyselimite2} requires: (1) an expression for the microscopic plastic dissipation $\pi(\textbf{d})$:
\begin{equation}
  \pi(\textbf{d}) = \sup2_{\sigma^{\star} \in C} \bm{\sigma}^{\star}:\textbf{d}
  \label{micropi}
  \end{equation}
and (2) a trial velocity field kinematically admissible with boundary conditions and verifying the property of incompressibility. Regarding the latter point, several trial velocity fields have been provided in previous studies \cite{thomason85a,thomasonnew1,thomasonnew2,hurebarrioz,keralavarma} for cylindrical porous unit-cell with cylindrical void, which have been shown to lead to good estimates of coalescence stress for both isotropic and (Hill)-anisotropic materials. For crystal plasticity as defined in Section~3.1.1, an analytical expression for the microscopic dissipation (Eq.~\ref{micropi}) is not available. An alternative is to use a regularized version of the crystal plasticity constitutive equations with the definition of a single yield criterion $\mathcal{F}$ through Arminjon-Gambin's equivalent stress:
\begin{equation}
  \mathcal{F} = \left[ \sum_k \left( \bm{\sigma}:\bm{\mu}^k \right)^n \right]^{1/n} - \tau_0
  \label{reguf}
  \end{equation}
An analytical expression for the microscopic plastic dissipation $\pi(\textbf{d})$ with the regularized model (Eq.~\ref{reguf}) is only known for $n=2$, which corresponds to a quadratic (Hill-type) approximation of the yield criterion \cite{paux}. However, it is known from previous studies that, in general, such approximation leads to poor results for the macroscopic plastic strength domain for single (porous) crystals. For the specific case considered here of evaluating the coalescence stress of porous single crystals, it has been checked in Appendix~A that such quadratic approximation does not lead to quantitative estimates of the coalescence stress, albeit capturing qualitatively the influence of crystallographic orientation. Therefore, a semi-analytical approach is followed in this study to estimate the microscopic dissipation:
\begin{equation}
  \pi(\textbf{d}) = \sum_k \tau_0\dot{\gamma}_k(\textbf{d}) = M(\textbf{d}) \tau_0 d_{eq}
  \label{micropim}
\end{equation}
where $ M(\textbf{d})$ is the Taylor factor defined by Eq.~\ref{taylorfactor}. Note that a similar approach has been used in \cite{nguyen}. Taylor factor can be easily calculated for arbitrary strain rate tensor and sets of slip systems using Taylor's microscopic minimum shear principle using a Simplex algorithm \cite{houtte88}. 

\subsubsection{Connection to void growth models}

As a case study, Eqs.~\ref{eqanalyselimite}, \ref{eqanalyselimite2} and \ref{micropim} are used for the homogenization of porous single crystals for diffuse plastic flow \cite{xuhan,paux}, to be compared to previous models. Considering a porous spherical unit-cell $\Omega$ (of radius $b$) with a spherical void $\omega$ (of radius $a$) under macroscopic hydrostatic loading (in spherical coordinates):
\begin{equation}
  \textbf{D} = D_{rr} \underline{e}_r \otimes \underline{e}_r
  \end{equation}
the classical trial velocity field (corresponding to the exact solution for an isotropic material) is:
\begin{equation}
  \textbf{d} = \frac{b^3}{3r^3}D_{kk} \left(-2 \textbf{e}_r \otimes  \textbf{e}_r + \textbf{e}_{\theta} \otimes  \textbf{e}_{\theta} + \textbf{e}_{\phi} \otimes  \textbf{e}_{\phi}    \right)
  \label{dhydro}
\end{equation}
The macroscopic dissipation can be approximated as:
\begin{equation}
  \begin{aligned}
    \Pi(\textbf{D}) \leq \,<M(\underline{r},\textbf{d})\tau_0 d_{eq} >_{\Omega} &= (1-f)<M(\underline{r},\textbf{d})\tau_0 d_{eq}>_{\Omega \backslash \omega} \\
                    &\approx <M(\underline{r},\textbf{d})>_{\Omega \backslash \omega} \tau_0 (1-f)<d_{eq}>_{\Omega \backslash \omega} 
\end{aligned}
\label{growthapp}
\end{equation}
where $\Omega \backslash \omega$ is the material around the void, and $f = (a/b)^3$ the porosity. Eq.~\ref{growthapp} is formally identical to the one used in Gurson model \cite{gurson} when the yield stress $\sigma_0$ is replaced by $<M(\underline{r},\textbf{d})>_{\Omega \backslash \omega} \tau_0$, leading to the hydrostatic point:
\begin{equation}
2f \cosh{\left(\frac{3 \Sigma_m}{2 <M(\underline{r},\textbf{d})>_{\Omega \backslash \omega} \tau_0}    \right)} - 1 - f^2 = 0 \ \ \ \ \ \Leftrightarrow \ \ \ \ \ \frac{\Sigma_m}{\tau_0} = -\frac{2}{3} <M(\underline{r},\textbf{d})>_{\Omega \backslash \omega} \ln{f}
\label{growthapp2}
\end{equation}
with $\Sigma_m = \Sigma_{kk}/3$ the macroscopic mean stress. Eq.~\ref{growthapp2} relies on the computation on the average Taylor factor:
\begin{equation}
    <M(\underline{r},\textbf{d})>_{\Omega \backslash \omega} = \frac{1}{vol \Omega \backslash \omega} \int_{\Omega \backslash \omega} M(\underline{r},\textbf{d})\,d\Omega
  \end{equation}
which can be interpreted as the average over random crystallographic orientations (through $\underline{r}$) of the Taylor factor for the strain-rate given by Eq.~\ref{dhydro}. This is exactly what is computed for the average Taylor factor of a polycrystalline aggregate without texture under uniaxial tension assuming Taylor iso-strain model \cite{taylor38}. For a FCC material:
\begin{equation}
  <M(\underline{r},\textbf{d})>_{\Omega \backslash \omega} = 3.066
  \label{MFCC}
  \end{equation}
This leads to a prefactor $\kappa$ before the normalized mean stress $\Sigma_m/\tau_0$ in Eq.~\ref{growthapp2} equals to $\kappa=0.489$, in close agreement with the prefactors obtained in previous studies: $\kappa = 0.513$ in \cite{xuhan}, $\kappa = 0.536$ in \cite{joessel} and $\kappa = 0.506$ in \cite{paux}, respectively through comparisons to numerical unit-cells results, based on infinite-rank laminates theory and computing directly the microscopic dissipation on Bishop-Hill vertices in the latter, thus leading to an exact value (for the microscopic strain-rate considered in Eq.~\ref{dhydro}). For this specific case of spherical porous unit-cell under hydrostatic loading conditions, the close agreement between the value found in \cite{paux} and the one obtained here (Eqs.~\ref{growthapp2} and \ref{MFCC}) shows that the approximation made in Eq.~\ref{growthapp} is relatively minor. A similar methodology can thus be applied to derive a coalescence criterion for porous crystals.

\subsection{Coalescence criterion for porous crystals}

A cylindrical unit-cell $\Omega$ of half-height $H$, radius $L$ with a coaxial cylindrical void $\omega$ of half-height $h$ and radius $R$ is considered for the derivation of the coalescence criterion for porous crystals. \textcolor{black}{This geometry corresponds to an approximation of a periodic hexagonal lattice of voids, and has been used previously in several studies \cite{thomasonnew1,thomasonnew2}}. As described in Section.~2, coalescence corresponds to a localized plastic flow in intervoid ligaments associated with elastic unloading above and below the voids, assuming rigid in the modeling.
\begin{figure}[H]
\includegraphics[height = 6cm]{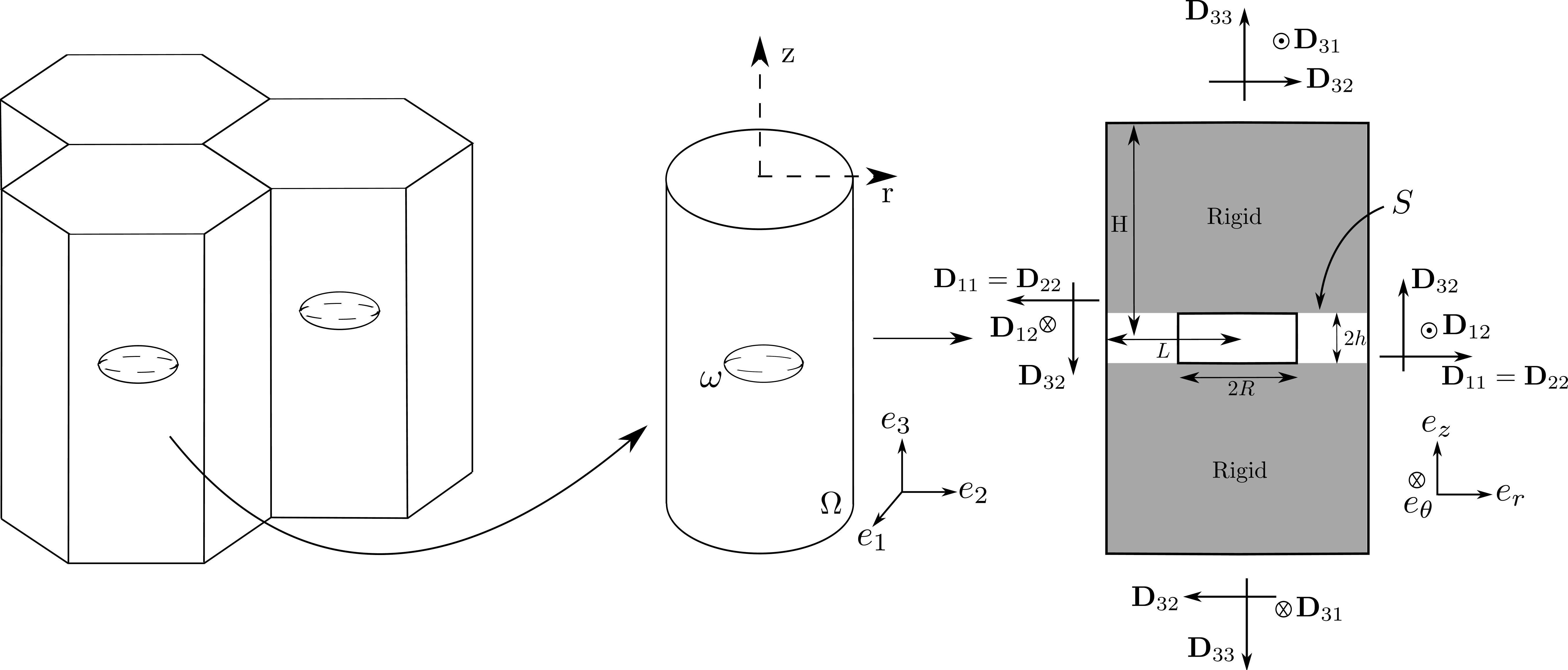}
\caption{Cylindrical unit-cell considered as an approximation of a unit-cell of a periodic array of voids of hexagonal lattice under periodic boundary conditions}
\label{figsk}
\end{figure}
\noindent
Two dimensionless ratios are used in the following, corresponding to the void aspect ratio and the intervoid ligament, respectively:
\begin{equation}
  W = \frac{h}{R} \ \ \ \ \ \ \ \ \ \ \ \ \ \ \ \chi = \frac{R}{L}
\end{equation}

\noindent
Two approximations are made:
\begin{itemize}
\item The velocity field associated with plastic flow in the intervoid ligament for single crystals can be approximated by the trial velocity field used for isotropic and (Hill-type) anisotropic matrix materials in previous studies;
\end{itemize}
\noindent
Such assumption is consistent with the results of Yerra \textit{et al.} \cite{yerra}, where the use of Thomason criterion (derived for isotropic materials) with an effective yield stress leads to a satisfactory agreement with numerical results, thus implying somehow that plastic flow at coalescence for isotropic materials and single crystals might be rather close.
\begin{itemize}
  \item Macroscopic plastic dissipation can be approximated through the multiplicative decomposition of average Taylor factor and average equivalent strain rate, as done in Eq.~\ref{growthapp}.
  \end{itemize}
The second approximation leads to a macroscopic plastic dissipation that takes formally the form $\Pi(\textbf{D}) = (1-f)\sigma_0 <d_{eq}>_{\Omega \backslash \omega}$ with $\sigma_0 = <M(\underline{r},\textbf{d})>_{\Omega \backslash \omega} \tau_0$. Thus, analytical coalescence criterion derived in previous studies for isotropic matrix materials can be used to describe porous single crystals by replacing the isotropic yield stress by $<M(\underline{r},\textbf{d})>_{\Omega \backslash \omega} \tau_0$. Such approximation appears to be similar in nature to the one proposed by Yerra \textit{et al.} \cite{yerra} (to compute an effective yield stress close to the void surface assuming equibiaxial strain state), but leads to a better description of the coupling between the crystallographic orientation and the non-uniform strain state in the intervoid ligament through the average Taylor factor. Using the trial velocity field proposed in \cite{thomasonnew1}, the coalescence criterion for porous single crystals under uniaxial straining conditions ($\textbf{D} = D_{33}\underline{e}_3 \otimes \underline{e}_3$) with respect to the coalescence plane takes the simple expression:
\begin{equation}
\textcolor{black}{\frac{\Sigma_{33}}{\tau_0} = M_1 t(W,\chi) \left[\frac{\chi^3 - 3\chi + 2}{3\sqrt{3} W\chi} \right]     + M_2\frac{b}{\sqrt{3}}\left[2 - \sqrt{1+3\chi^4} + \ln \frac{1 + \sqrt{1 + 3\chi^4}}{3\chi^2}    \right]}
\label{eqtorki}
\end{equation}
where $M_1$ and $M_2$ are averages of Taylor factors, $b=0.9$ and $t(W,\chi) = [W(-0.84+12.9\chi)]/[1+W(-0.84+12.9\chi)]$ two fitting parameters that have been adjusted in \cite{torki} based on numerical data for von Mises isotropic material. The two terms in Eq.~\ref{eqtorki} have different origins: the first one corresponds to the tangential discontinuity of the velocity field at $|z|=h$, corresponding to a constant strain rate in an interphase of thickness $e$:
\begin{equation}
\textbf{d}_1 = d_{eq}^1(r) \frac{\left(\underline{e}_r \otimes \underline{e}_z + \underline{e}_z \otimes \underline{e}_r    \right)}{2/\sqrt{3}} = d_{eq}^1(r) \bar{\textbf{d}}_1
\end{equation}  
The average Taylor factor $M_1$ is computed in the interface where the tangential discontinuity lies\footnote{or equivalently in the interphase of thickness $e$, as the strain rate is constant.}:
\begin{equation}
  \begin{aligned}
    M_1 = < M(\underline{r},\textbf{d}_1)>_{S} = < M(\underline{r},\bar{\textbf{d}}_1)>_{S} = &= \frac{1}{\pi(L^2 - R^2)} \int_{r=R}^L \int_{\theta=0}^{2\pi} M(r,\theta,\bar{\textbf{d}}_1) r dr d\theta\\
    &= \frac{1}{\pi(1 - \chi^2)} \int_{r=\chi}^1 \int_{\theta=0}^{2\pi} M(\bar{r},\theta,\bar{\textbf{d}}_1) \bar{r} d\bar{r} d\theta
  \end{aligned}
  \label{eqM1}
  \end{equation}
The second term in Eq.~\ref{eqtorki} comes from the plastic dissipation in the intervoid ligament where the strain rate is:
\begin{equation}
  \textbf{d}_2 = d_{eq}^2(\bar{r}) \left( \frac{- 1 - \bar{r}^{-2}}{2\sqrt{(3 + \bar{r}^{-4})/3}}\underline{e}_r \otimes \underline{e}_r + \frac{-1 + \bar{r}^{-2}}{2\sqrt{(3 + \bar{r}^{-4})/3}}\underline{e}_{\theta} \otimes \underline{e}_{\theta} + \frac{1}{\sqrt{(3 + \bar{r}^{-4})/3}}\underline{e}_z \otimes \underline{e}_z \right) = d_{eq}^2(\bar{r})\bar{\textbf{d}}_2
  \label{secondterm}
  \end{equation}
with ($\bar{r} = r/L$). The average Taylor factor $M_2$ is computed over the intervoid ligament:
\begin{equation}
  \begin{aligned}
    M_2 = < M(\underline{r},\textbf{d}_2)>_{\Omega \backslash \omega}= < M(\underline{r},\bar{\textbf{d}}_2)>_{\Omega \backslash \omega} &= \frac{1}{\pi h (1 - \chi^2)} \int_{r=\chi}^1 \int_{\theta=0}^{2\pi} \int_{z=0}^{h} M(r,\theta,\bar{\textbf{d}}_2) \bar{r} d\bar{r} d\theta dz\\
    &= \frac{1}{\pi(1 - \chi^2)} \int_{r=\chi}^1 \int_{\theta=0}^{2\pi} M(\bar{r},\theta,\bar{\textbf{d}}_2) \bar{r} d\bar{r} d\theta
  \end{aligned}
  \label{eqM2}
  \end{equation}
The coalescence criterion (Eq.~\ref{eqtorki}) is valid in absence of shear with respect to the coalescence plane. Similarly, in presence of shear $\Sigma_{sh}$ associated with a macroscopic strain:
\begin{equation}
  \textbf{D} = D_{33} \underline{e}_3 \otimes \underline{e}_3 + D_{sh}\left(\underline{e}_s \otimes \underline{e}_h + \underline{e}_h \otimes \underline{e}_s  \right) \ \ \ \ \ \mathrm{with} \ \ \ \ \ \{\underline{e}_s,\underline{e}_h\} \in \{\underline{e}_1,\underline{e}_2 \}
  \end{equation}
the coalescence criterion proposed in \cite{torki} can be extended for porous single crystals as:
\begin{equation}
\left\{  
\begin{aligned}
\textcolor{black}{\frac{(|\Sigma_{33}| - t(W,\chi)\Sigma^{surf} )^2}{b^2\Sigma^{vol^2}} + 4 \frac{\Sigma_{sh}^2}{T^2} - 1} & \textcolor{black}{= 0} & \textcolor{black}{\mathrm{for}\ |\Sigma_{33}| \geq \Sigma^{surf}}\\ 
\textcolor{black}{4 \frac{\Sigma_{sh}^2 }{T^2} - 1} & \textcolor{black}{= 0} & \textcolor{black}{\mathrm{for}\ |\Sigma_{33}| \leq \Sigma^{surf}}\\ 
\end{aligned}
\right.
\label{eqfullshear}
\end{equation}
where $T$ is the coalescence stress under pure shear:
\begin{equation}
  \textcolor{black}{\frac{T}{\tau_0}} \textcolor{black}{=\frac{2 M_3}{\sqrt{3}} \left(1 - \chi^2   \right)}
  \label{eqfullshear2}
\end{equation}
The average Taylor factor $M_3$ is computed over the intervoid ligament in which the trial velocity field chosen in \cite{torki} leads to a constant microscopic strain-rate tensor $\textbf{d}_3 = D_{sh} \left(\underline{e}_s \otimes \underline{e}_h + \underline{e}_h \otimes \underline{e}_s  \right) $:
\begin{equation}
  \begin{aligned}
    M_3 = < M(\underline{r},\textbf{d}_3)>_{\Omega \backslash \omega} &= \frac{1}{\pi h (L^2 - R^2)} \int_{r=R}^L \int_{\theta=0}^{2\pi} \int_{z=0}^{h} M(r,\theta,\textbf{d}_3) r dr d\theta dz\\
    &=  M(\textbf{d}_3) 
  \end{aligned}
  \label{eqM3}
  \end{equation}
The set of equations~\ref{eqtorki},~\ref{eqM1},~\ref{eqM2},~\ref{eqfullshear},~\ref{eqfullshear2},~\ref{eqM3} gives a coalescence criterion for porous single crystals under arbitrary loading conditions. This criterion is validated against numerical data in the next section. Average Taylor factors (Eqs.~\ref{eqM1},~\ref{eqM2},~\ref{eqM3}) are computed by adaptative quadrature scheme where the integrand function (local Taylor factor) relies on Taylor's minimum shear principle solved through linear programming \cite{mtex}.

\section{Validation of the coalescence criterion}

\subsection{Numerical limit-analysis}

The coalescence criterion derived in Section~3 is assessed by comparisons to numerical limit-analysis simulations carried out to get the plastic strength domain of porous single crystals subjected to boundary conditions relevant for coalescence, consistently with the hypothesis used in the theoretical derivation. \texttt{AMITEX\_FFTP} Fast-Fourier-Transform (FFT) based solver has been used. \textcolor{black}{This numerical method, introduced in \cite{moulinec}, allows to solve the mechanical equilibrium equations of periodic unit cells with an iterative algorithm based on the Lippman-Schwinger equation and discrete Green operator. \texttt{AMITEX\_FFTP} solver \cite{amitex} uses fixed point scheme proposed in \cite{moulinec}, but with filtered discrete Green operator \cite{gelebartouaki} and convergence acceleration \cite{ramiere} that both contribute to improving the convergence and to reduce the sensibility to material contrast (differences in local stiffness) encountered in the original implementation.} Small strain assumption is used, with crystal plasticity constitutive equations for the matrix material. A regularized version of the model described in Section~3 is used where the plastic flow rule reads:
\begin{equation}
  \dot{\gamma}_k = \left\langle \frac{|\tau_k| - \tau_0}{K} \right\rangle^n sgn(\tau_k)
  \end{equation}
Isotropic elasticity has been used (through Hooke's law and additive decomposition of the total strain into elastic strain and plastic strain), but does not affect the plastic strength domain. Constitutive equations have been implemented in the \texttt{MFront} code generator \cite{mfront}, using a implicit discretization solved with a Newton-Raphson algorithm. Details of the numerical implementation can be found in \cite{hure2016}. Young's modulus and Poisson ratio have been set arbitrarily to $Y=200$GPa and $\nu=0.3$. Parameters $\tau_0$, $K$ and $n$ have been set to $100$MPa, $10$MPa and $15$, respectively. These parameter values, already used in \cite{xuhan}, ensure a negligible strain rate dependency for all simulations performed. In addition, simulations have also been performed assuming isotropic von Mises plasticity for the matrix material (with yield stress $\tau_0$) as a reference case. 
\begin{figure}[H]
\includegraphics[height = 5cm]{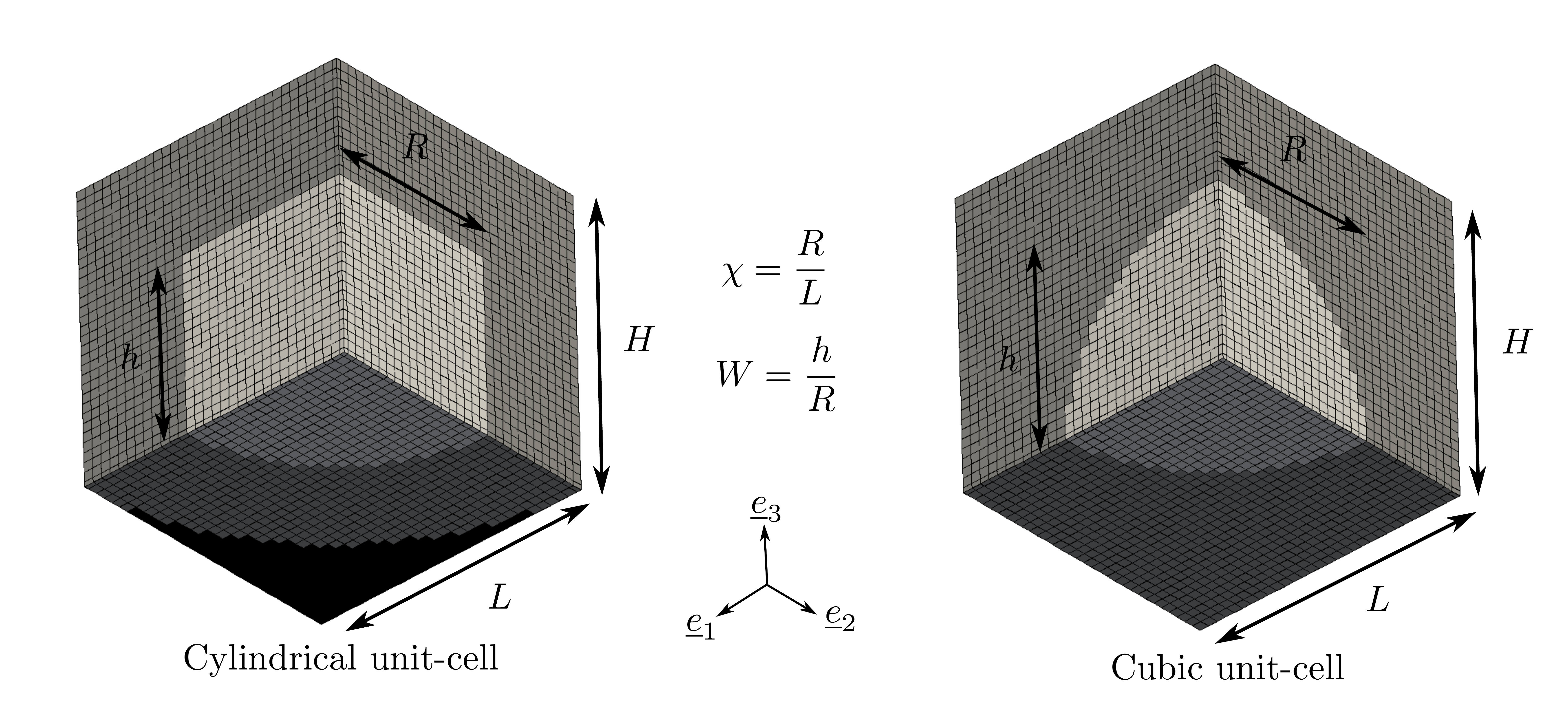}
\caption{Unit-cells used for FFT simulations. Constitutive equations are assigned to each voxel: dark gray corresponds to crystal (or von Mises) plasticity, light gray to an elastic material with zero stiffness and black to an elastic material with Young's modulus $Y_{33}$ and shear moduli $G_{31},\ G_{31}$ equal to zero, and (almost-)infinite Young's moduli $Y_{11},\ Y_{22}$ and shear modulus $G_{12}$.}
\label{figuc}
\end{figure}
\noindent
Two different unit-cells are used (Fig.~\ref{figuc}). The first one aims at simulating a cylindrical unit-cell with a cylindrical void, consistently with the geometry considered in Section~3. As FFT simulations relie on periodic grids where each voxel is assigned a given set of constitutive equations, a fictive material is used around the cylindrical unit-cell to ensure that  under uniaxial straining conditions prevail. This can be done through an elastic material with Young's modulus $Y_{33}$ and shear moduli $G_{31},\ G_{32}$ equal to zero, and (almost)-infinite Young's moduli $Y_{11},\ Y_{22}$ and shear modulus $G_{12}$. Details can be found in \cite{barriozJAM}. Crystal plasticity (or von Mises plasticity) is assigned to the voxels inside the cylindrical unit-cell, and elastic material with zero stiffness is set to the voxels for the cylindrical void. This unit-cell is used only to study loading in absence of shear. A second kind of unit-cell is used corresponding to a simple cubic arrangement of spheroidal voids.\\

\noindent
Constant macroscopic strain-rate corresponding to coalescence are applied:
\begin{equation}
  \textbf{D} =  D_{33} \underline{e}_3 \otimes \underline{e}_3 + \alpha D_{33} \left(  \underline{e}_1 \otimes \underline{e}_3 +  \underline{e}_3 \otimes \underline{e}_1 \right)
  \label{loading}
  \end{equation}
with $D_{33} =10^{-5} \mathrm{s^{-1}}$, and $\alpha$ a parameter allowing to assess the effect of shear stress on coalescence criterion. Simulations are run until saturation of the macroscopic stresses which are then compared to the coalescence criterion proposed in Section~3. A convergence study with respect to the number of voxels has been performed, and the error associated with each result presented hereafter is estimated to be less than 1\%. Simulations are performed for different void aspect ratio $W = [0.5;1;3]$ and intervoid ligament $\chi \in [0.3:0.7]$. Two sets of slip systems are considered: Face-Centered Cubic (FCC) and Hexagonal Close Packed (HCP) with only second order Pyramidal slip system (Tab.~\ref{tabslip}). The latter is not intended to be realistic, but is rather used as a test of strong plastic anisotropy compared to FCC.

\begin{table}[H]
\begin{tabular}{c|c|c}
Crystal system &  \multicolumn{1}{c}{Slip plane $\underline{n}$\ / direction $\underline{m}$}  &  Number of slip systems \\
\hline
\hline
FCC          &  $\{111\}$ $\langle 110 \rangle$    &  12 \\
\hline
HCP         &  $\{11\bar{2}2\}$ $\langle 11\bar{2}{\bar{3}} \rangle$     &   6  \\
\hline
\end{tabular}
\caption{Crystal systems and associated slip systems considered in this study}
\label{tabslip}
\end{table}
Several crystallographic orientations have been considered (similar to the ones used in \cite{xuhan,chaoling} for FCC), which are defined in Tab.~2 and visualized in Fig.~6.

\begin{figure}[H]
  \begin{floatrow}
    \capbtabbox{
  \begin{tabular}{c|c|c|c|c}
    Crystal system & Reference & \multicolumn{3}{c}{Orientation} \\
    &  &$e_3$ & $e_2$ & $e_1$ \\
       \hline
       \hline
       FCC & FCC100 & [100]  & [001]   & [010]     \\
       FCC & FCC110 & [110]  & [001]   & [-110]     \\
       FCC & FCC111 & [111]  & [0-11]   & [-211]     \\
       FCC & FCC210 & [210]  & [001]   & [-120]     \\
       FCC & FCC-125 & [-125]  & [210]  & [1-21]     \\
       \hline
       HCP & HCP1& [10-10]  & [0001]    & [-12-10]     \\
       HCP & HCP2& [-12-10]  & [0001]   & [10-10]     \\
       HCP & HCP3& [0001]   & [10-10]   &  [-12-10]    \\
       \hline
       \label{tabori}
  \end{tabular}
  }
        {
                 \caption{Definition of the crystallographic orientations}
        }
        \hspace{-1cm}
\ffigbox{               
  \includegraphics[height = 3.cm]{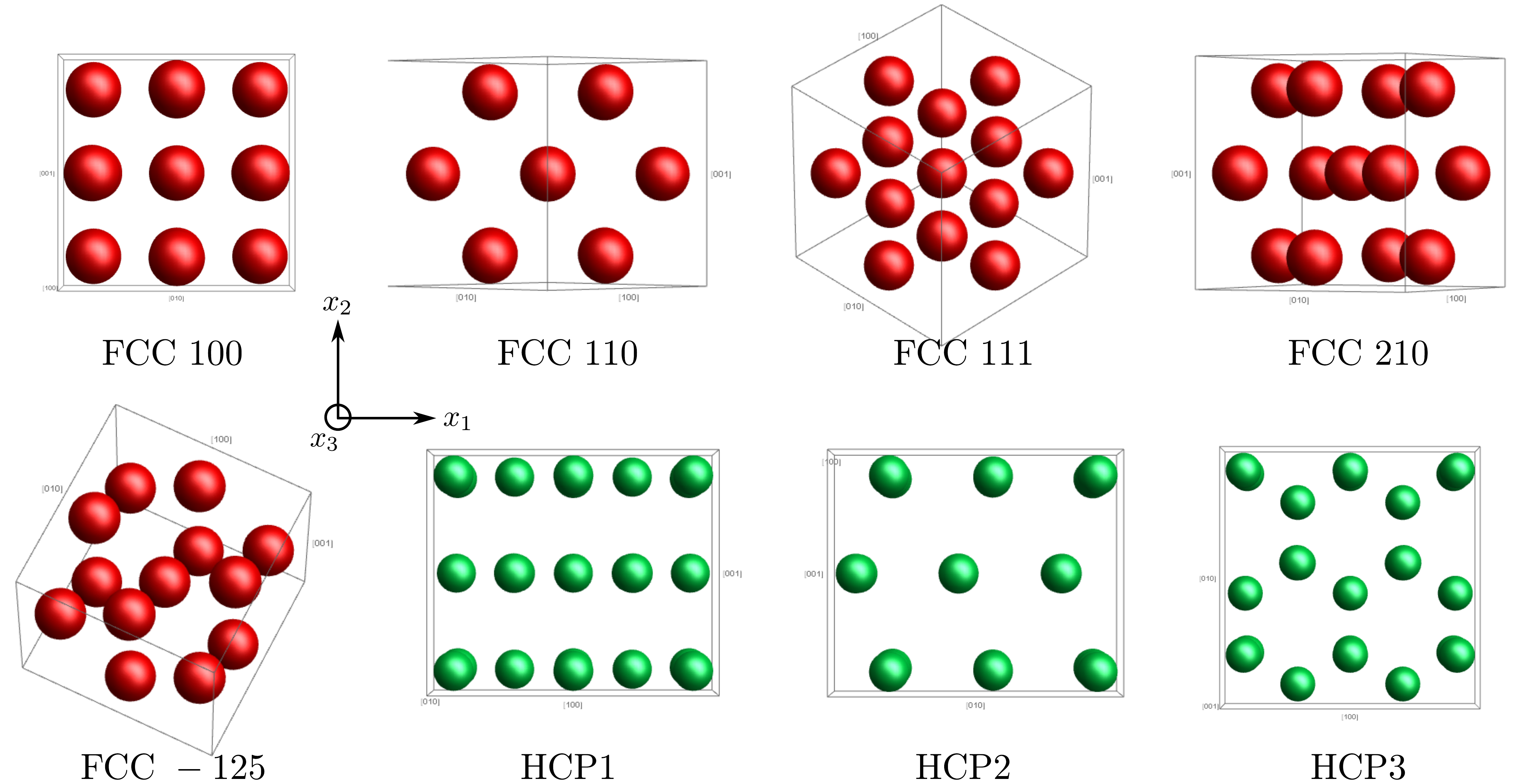}
}
{\caption{Visualization of the atom lattices for the different crystallographic orientations}}
  \end{floatrow}
\end{figure}
\noindent
Equivalent strain rate $\dot{\epsilon}_{eq}=\sqrt{(2/3)\bm{\dot{\epsilon}}:\bm{\dot{\epsilon}}}$ fields are computed after each simulation
to be compared to plastic strain-rate field $d_{eq}$ used in Section~3.

\subsection{Uniaxial straining conditions}
\subsubsection{Cylindrical voids / Cylindrical unit-cells}

Coalescence criterion predictions are compared to numerical results in the case of uniaxial straining conditions. Cylindrical voids in cylindrical unit-cells are firstly considered in the simulations, consistently with the geometry used in the derivation of the coalescence criterion. Results are presented on Fig.~\ref{figresu1}a,~b,c~ for various values of the void aspect ratio and intervoid ligament. As a reference case, von Mises perfect plasticity has been used for the matrix material. A very good agreement is observed between predictions (Eq.~\ref{eqtorki} with $M_1=M_2=1$) and numerical results for the range of parameters considered, in agreement with the results shown in \cite{torki}. This validates using periodic unit-cells and FFT-based solver to study cylindrical unit-cells under coalescence conditions. Reference numerical equivalent strain rate fields are also shown in the insets of Fig.~\ref{figresu1}b,~d,~f, to be compared to numerical equivalent strain rate fields for single crystals matrix materials. \textcolor{black}{In addition, the equivalent strain rate field corresponding to the trial velocity field (Eq.~\ref{secondterm}) is shown in the inset of Fig.~\ref{figresu1}d for $W=1,\chi=0.5$.}

\begin{figure}[H]
\centering
\subfigure[]{\includegraphics[height = 4.cm]{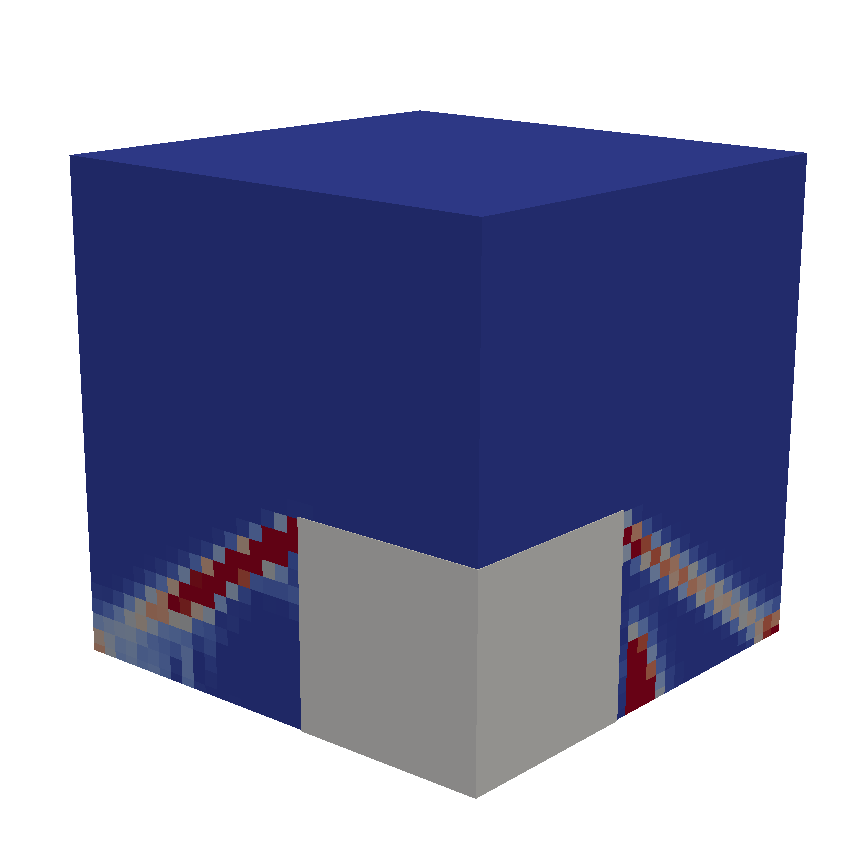}}
\hspace{2cm}
\subfigure[]{\includegraphics[height = 4.cm]{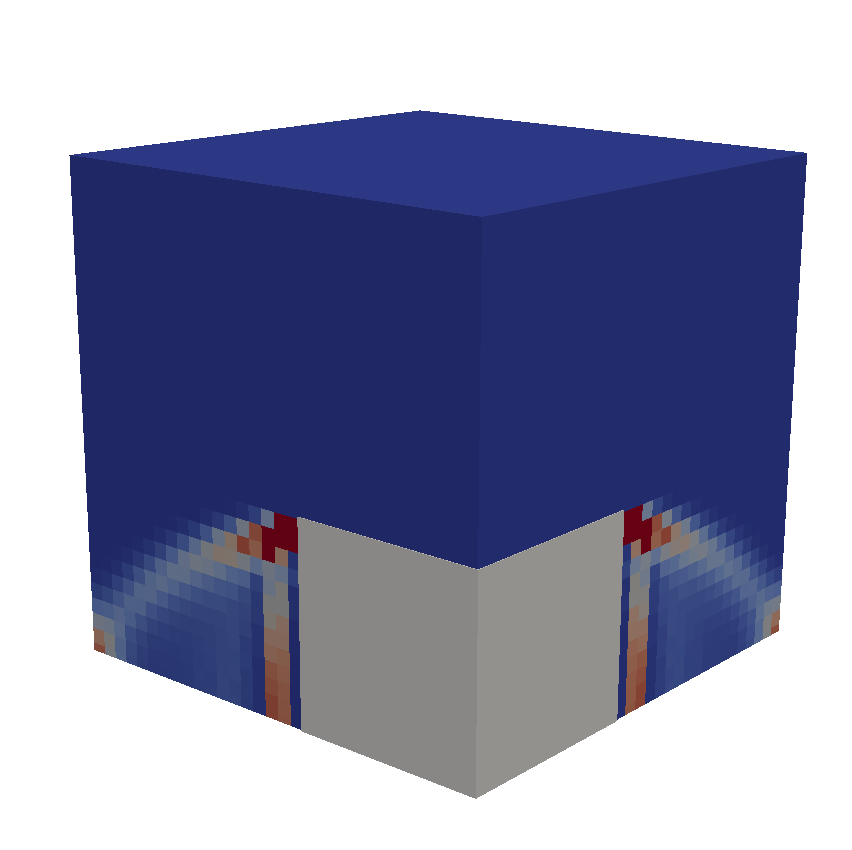}}
\caption{Numerical equivalent strain rate field (in arbitrary units) for cylindrical voids ($W=1$, $\chi=0.4$) in cylindrical unit-cells, for (a) HCP3 matrix material and (b) isotropic von Mises material}
\label{deqhcp}
\end{figure}

Normalized coalescence stresses for FCC single crystals as a function of intervoid ligament size for three different orientations ([100], [110] and [210]) are shown on Fig.~\ref{figresu1}a,~b,~c, as well as predictions using Eq.~\ref{eqtorki}. Additional simulations with orientations [111] and [-125], not shown here, have also been performed, and results lie in between the results for the other orientations. Numerical results indicate a weak effect of FCC crystal plastic anisotropy on the coalescence stress in uniaxial straining conditions. Such observation is consistent with results presented in \cite{yerra} (for BCC material) based on finite-strain porous unit-cells simulations, and is inferred to be related to the large number of slip systems for FCC/BCC materials. The equivalent strain rate fields shown on Fig.~\ref{figresu1}b,~d,~f clearly exhibit the traces of the slip systems marked by high values of strain rate along well-defined regions. However, it should be noted that the overall field is in qualitative agreement with the one observed for isotropic (Mises) material. 

\begin{figure}[H]
\centering
\subfigure[]{\includegraphics[height = 3.5cm]{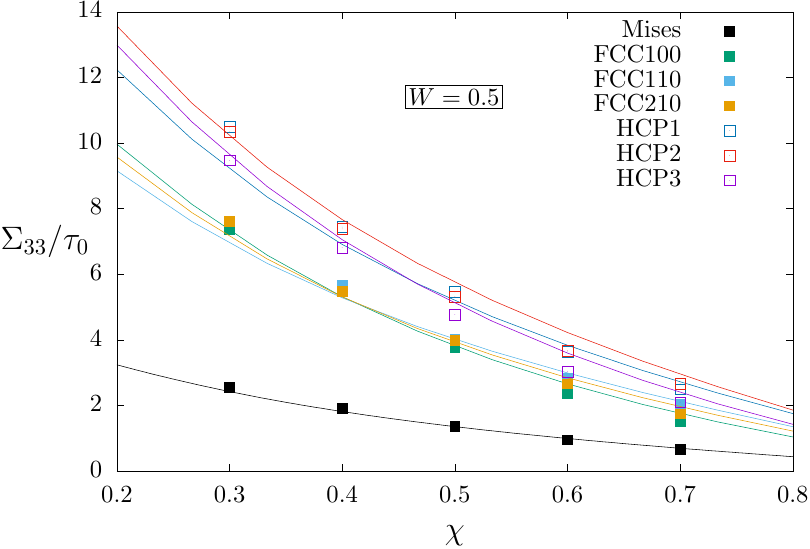}} \hspace{1.3cm}
\subfigure[]{\includegraphics[height = 3.5cm]{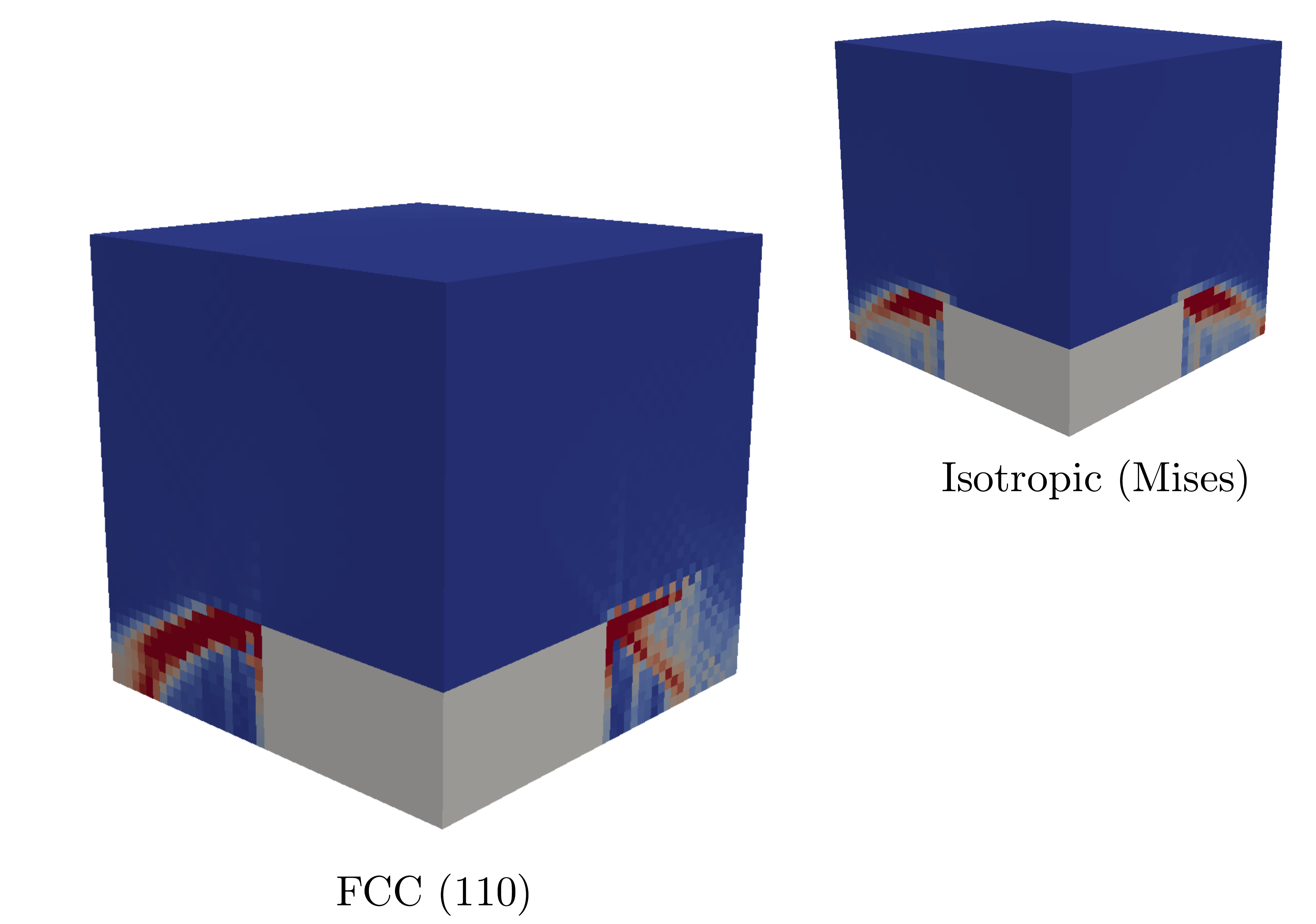}}
\subfigure[]{\includegraphics[height = 3.5cm]{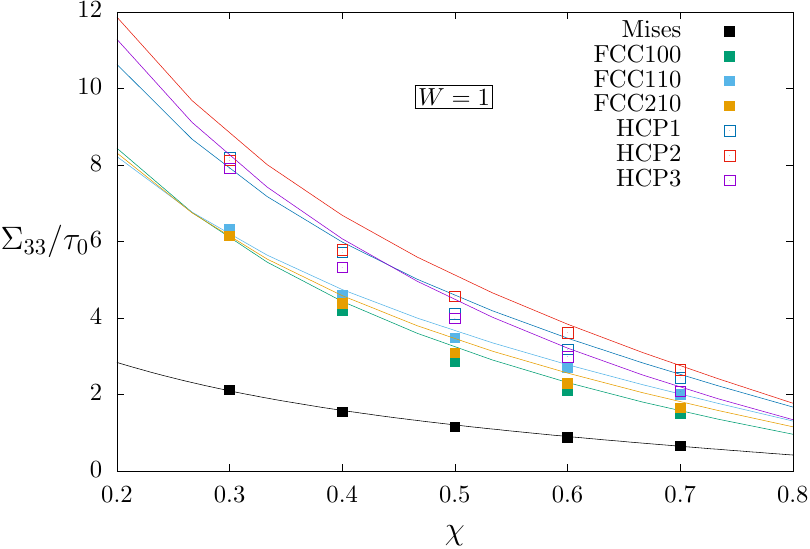}} \hspace{1.3cm}
\subfigure[]{\includegraphics[height = 3.5cm]{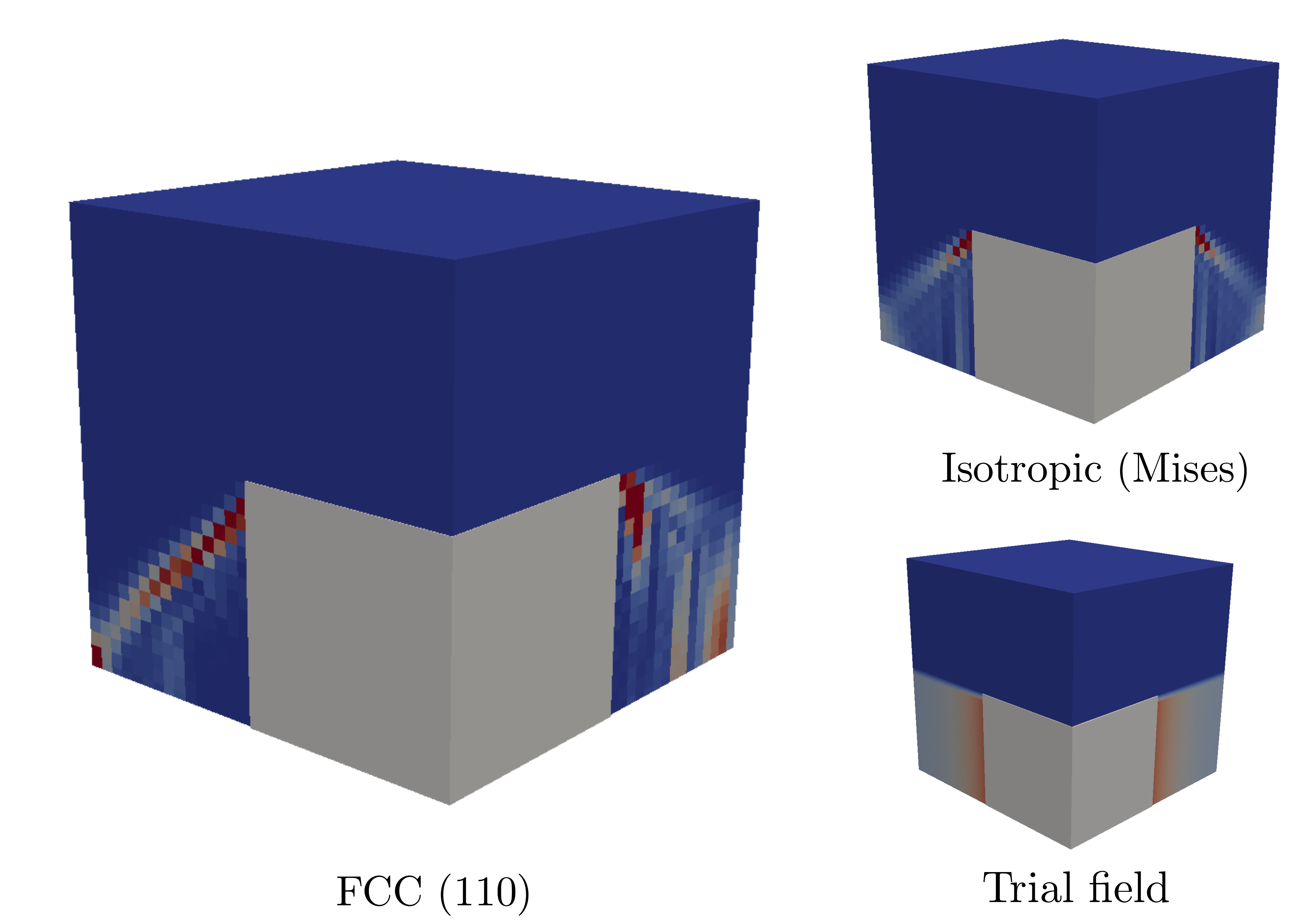}}
\subfigure[]{\includegraphics[height = 3.5cm]{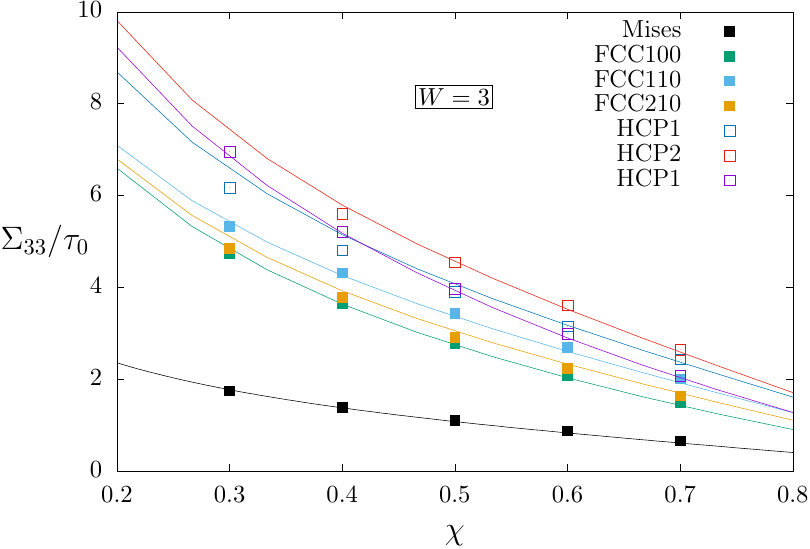}} \hspace{1.3cm}
\subfigure[]{\includegraphics[height = 3.5cm]{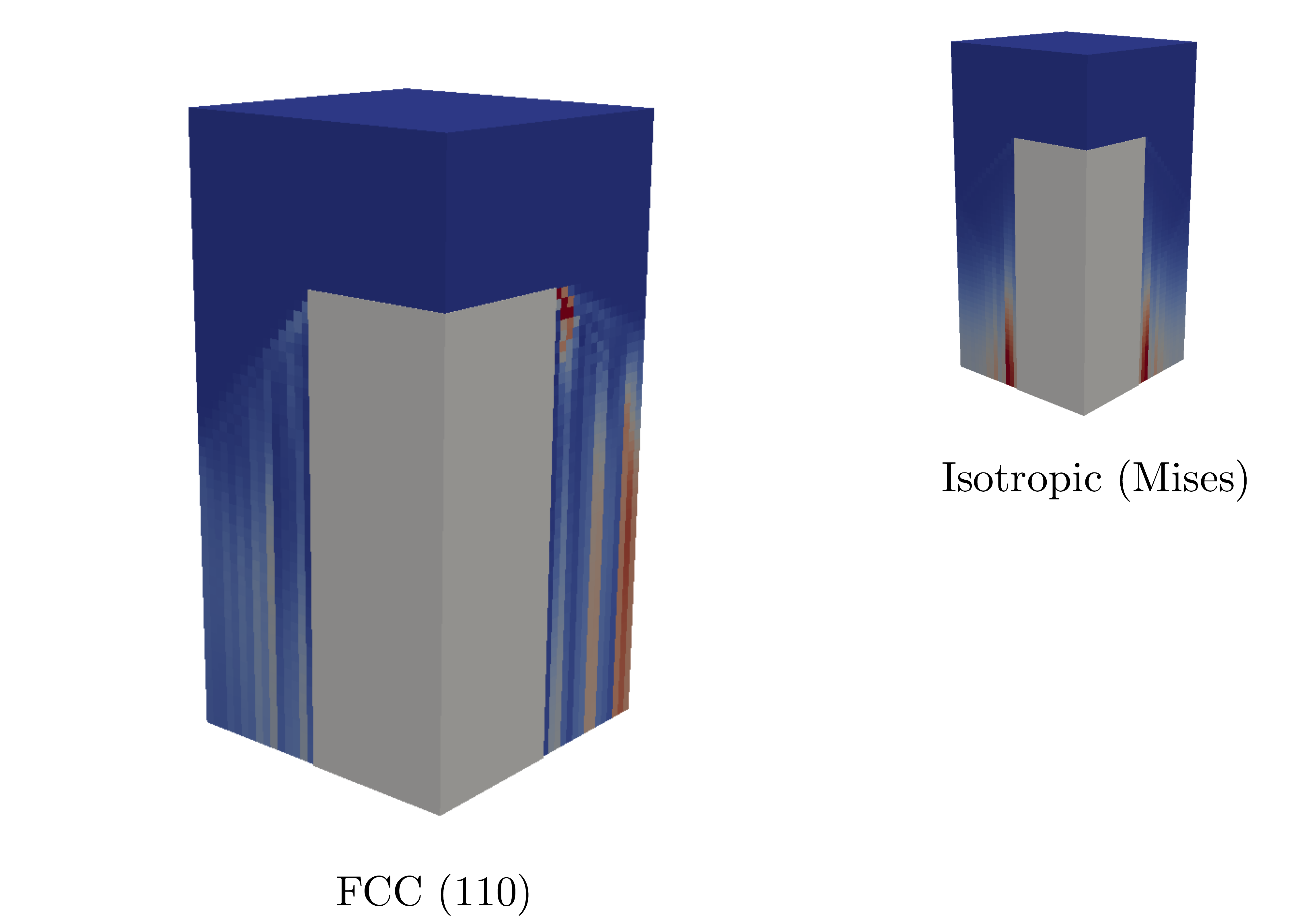}} \hspace{0.cm}
\caption{Normalised coalescence stress as a function of the dimensionless intervoid ligament size, for different constitutive equations  (von Mises, crystal plasticity (FCC, HCP) with different crystallographic orientations). Comparisons between predictions (Eq.~\ref{eqtorki}, solid lines) and numerical results (squares), for cylindrical voids in cylindrical unit-cells, for $W=0.5$ (a), $W=1$ (c) and $W=3$ (e). Associated numerical equivalent strain rate fields (in arbitrary units) for FCC material oriented along $[110]$ (b,~d,~f). Insets: Numerical equivalent strain rate fields for an isotropic von Mises material. \textcolor{black}{(d) Inset: Equivalent strain rate field corresponding to the trial velocity field (Eq.~\ref{secondterm})}}
\label{figresu1}
\end{figure}

\textcolor{black}{As shown on Fig.~\ref{figresu1}d, the trial velocity field is rather a crude approximation of the numerical results, even for the isotropic material. The tangential discontinuity of the trial field (leading also to a strain rate field at the interface with the rigid regions, not represented in Fig.~\ref{figresu1}d) is somehow able to capture what happens in that region, as far as coalescence stress is concerned.} As a whole, the predictions of the proposed coalescence criterion are in very good agreement with numerical results for FCC crystal, capturing quantitatively the weak effect of crystallographic orientation on coalescence stress, for both orientations with high and low symmetries ([100] \textit{vs.} [210]).

Normalized coalescence stresses for HCP single crystals with three different orientations are also shown on Fig.~\ref{figresu1}a,~b,~c. Numerical results are clearly higher, for similar geometrical parameters, than those obtained for FCC single crystals. This is attributed to the lower number of available slip systems for the HCP material considered in this study than for FCC\footnote{\textcolor{black}{It is reminded that the HCP material considered in this study, accounting only for second order pyramidal set of slip systems with identical critical resolved shear stresses (and no other slip systems nor twinning), is not intended to be realistic, but is used to assess the model for higher plastic anisotropy (compared to FCC).}}. The increase of coalescence stress of HCP materials is also quantitatively captured by the coalescence criterion for the different conditions studied. The effect of crystallographic orientation on coalescence stress for HCP material considered is well described by the coalescence criterion for large values of the intervoid ligament $\chi$. For lower values, discrepancies are observed, which are mainly attributed to the poor description of the numerical strain rate fields by their isotropic counterparts (Fig.~\ref{deqhcp}). In this particular case, a single slip system appear to be activated in a large part of the material (compared to somehow more diffuse fields for FCC110 as shown on Fig.~\ref{figresu1}d), which can not be captured by an isotropic strain rate field.

\begin{figure}[H]
\includegraphics[height = 8cm]{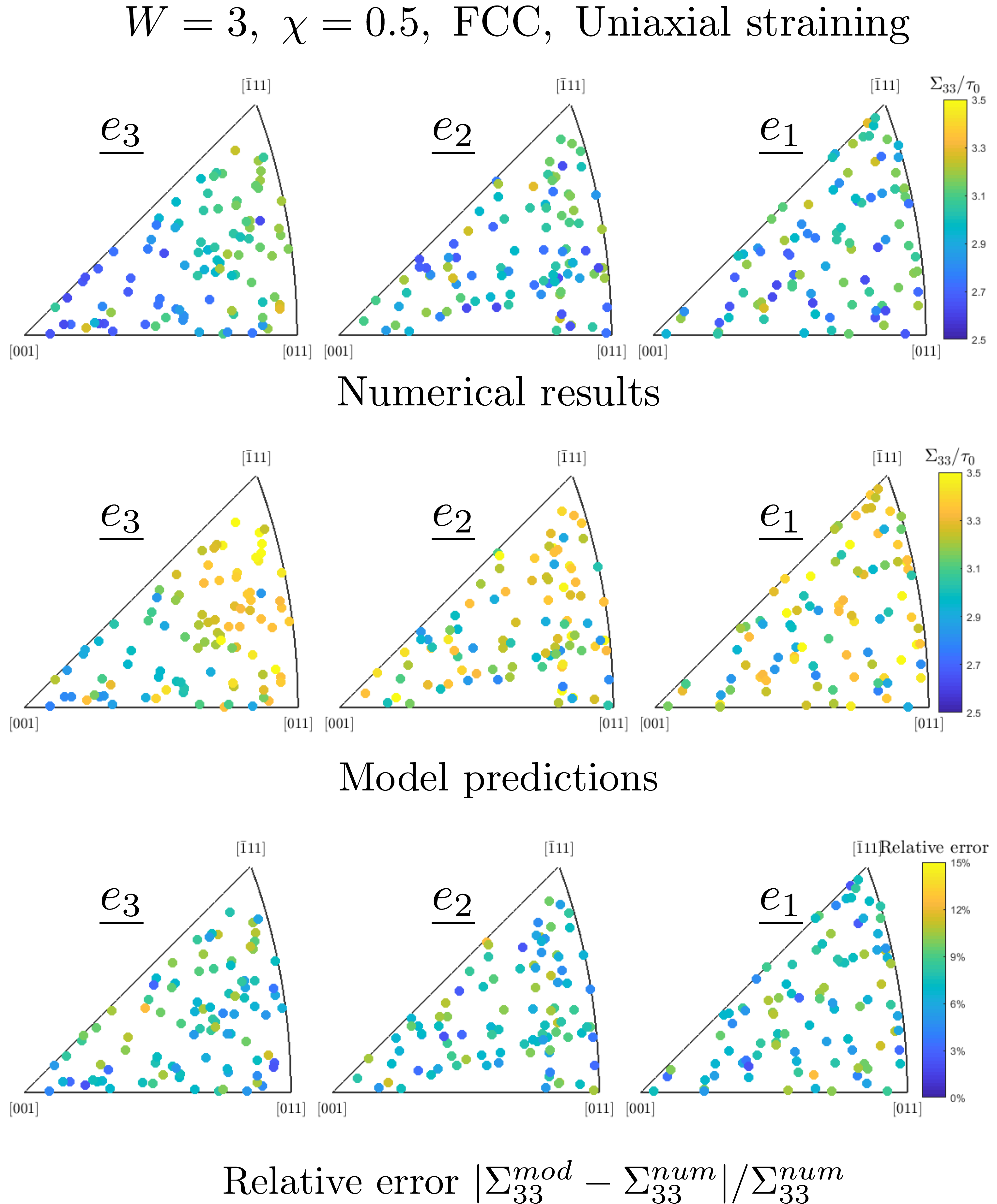}
\caption{\textcolor{black}{Normalised coalescence stress for $W=3$, $\chi=0.5$ for FCC material under uniaxial straining conditions, as a function of crystallographic orientations (represented in inverse pole figures). Comparisons between numerical results, model predictions and associated relative errors}}
\label{figerror}
\end{figure}

\textcolor{black}{The accuracy of the proposed criterion is found to be rather good for all orientations tested in Fig.~\ref{figresu1}, for cylindrical void in cylindrical unit-cell. In order to assess more in details the accuracy of the model with respect to the crystallographic orientation, additional simulations are performed for a specific void configuration ($W=3,\chi=0.5$) for FCC material. One hundred random crystallographic orientations are used, and coalescence stresses are computed using numerical limit analysis and using Eqs.~\ref{eqtorki},~\ref{eqM1},~\ref{eqM2}. Results are presented in Fig.~\ref{figerror} in inverse pole figures, as well as relative errors. For both numerical results and model predictions, a strong dependence of the coalescence stress under uniaxial straining conditions to the crystal orientation along the loading axis is found. Coalescence stress is higher when the crystal is oriented along the [111] axis, as such orientation corresponds to slip systems (of normal [111]) not well-oriented. Relative errors between numerical results and model predictions are found to be below $13\%$, with an average relative error of $7\%$. These results confirm the good accuracy of the proposed criterion to describe the effect of crystallographic orientation. A complete assessment of coalescence stress (and comparisons to model predictions) would however be required (as a function of the set of slip systems considered, void aspect ratio, intervoid distance), which is left for a future study}  

\subsubsection{Spheroidal voids / Cubic unit-cells}

Eqs.~\ref{eqtorki},~\ref{eqM1},~\ref{eqM2} have been found to be in good agreement with numerical results using the same geometry as the one considered in the theoretical derivation, corresponding to a cylindrical void in a cylindrical unit-cell. This validates both the use of an isotropic velocity field to describe plastic flow in porous single crystals under uniaxial straining conditions, and also the approximation made to compute the plastic dissipation. Predictions from Eqs.~\ref{eqtorki},~\ref{eqM1},~\ref{eqM2} are now compared to numerical results obtained for spheroidal voids in cubic unit-cells. 

\begin{figure}[H]
\centering
\subfigure[]{\includegraphics[height = 4.cm]{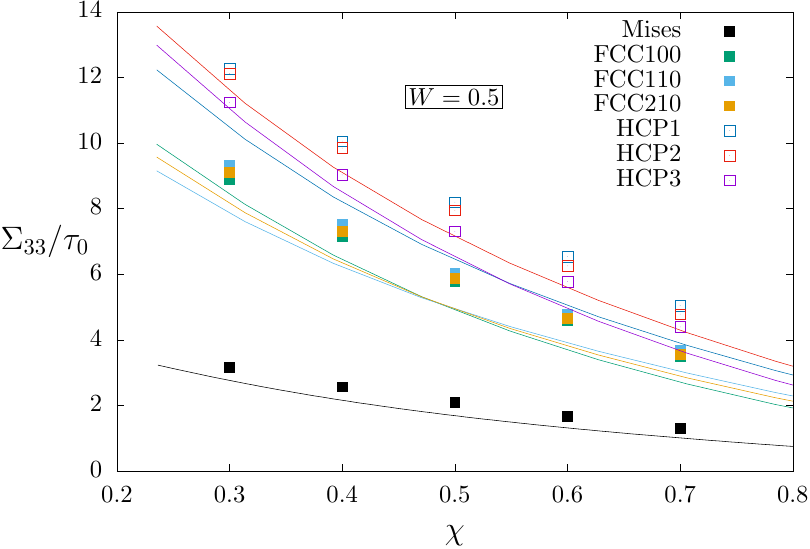}} \hspace{1.3cm}
\subfigure[]{\includegraphics[height = 4.cm]{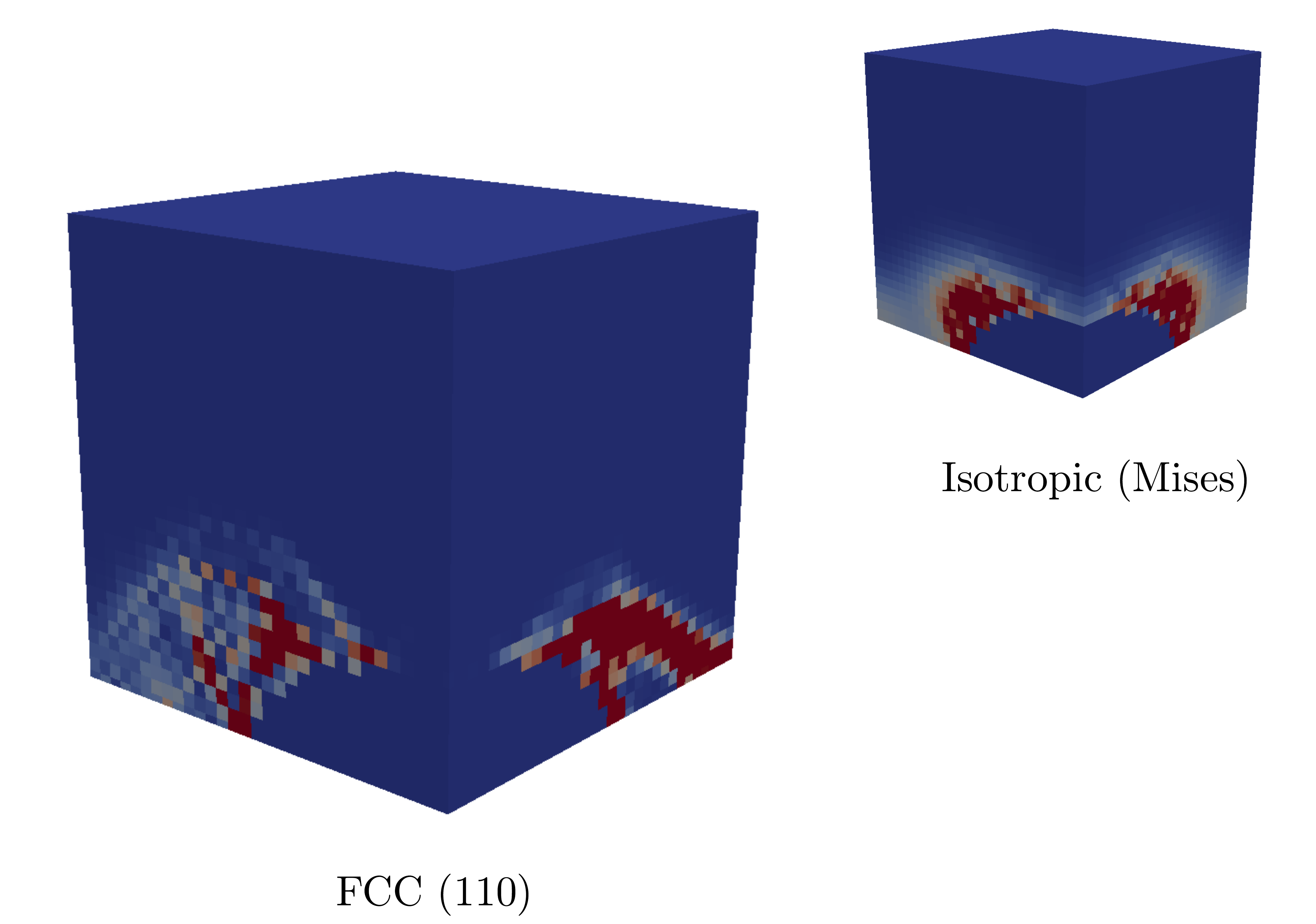}}
\subfigure[]{\includegraphics[height = 4.cm]{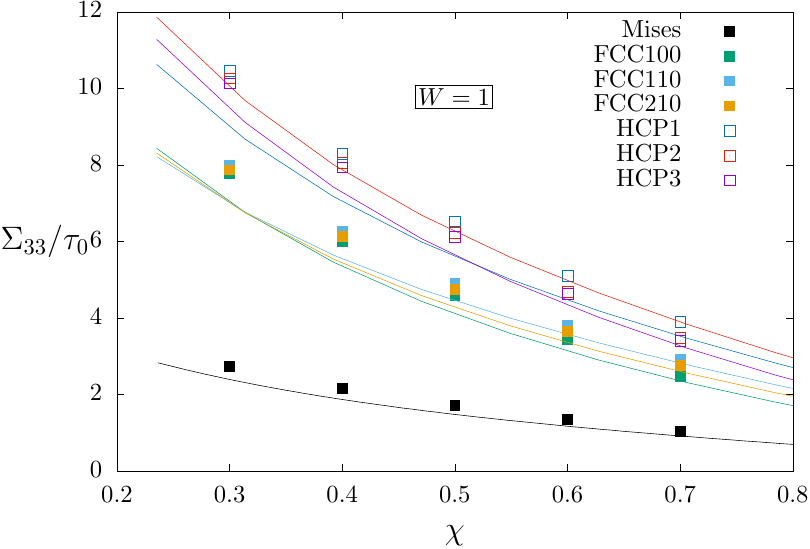}} \hspace{1.3cm}
\subfigure[]{\includegraphics[height = 4.cm]{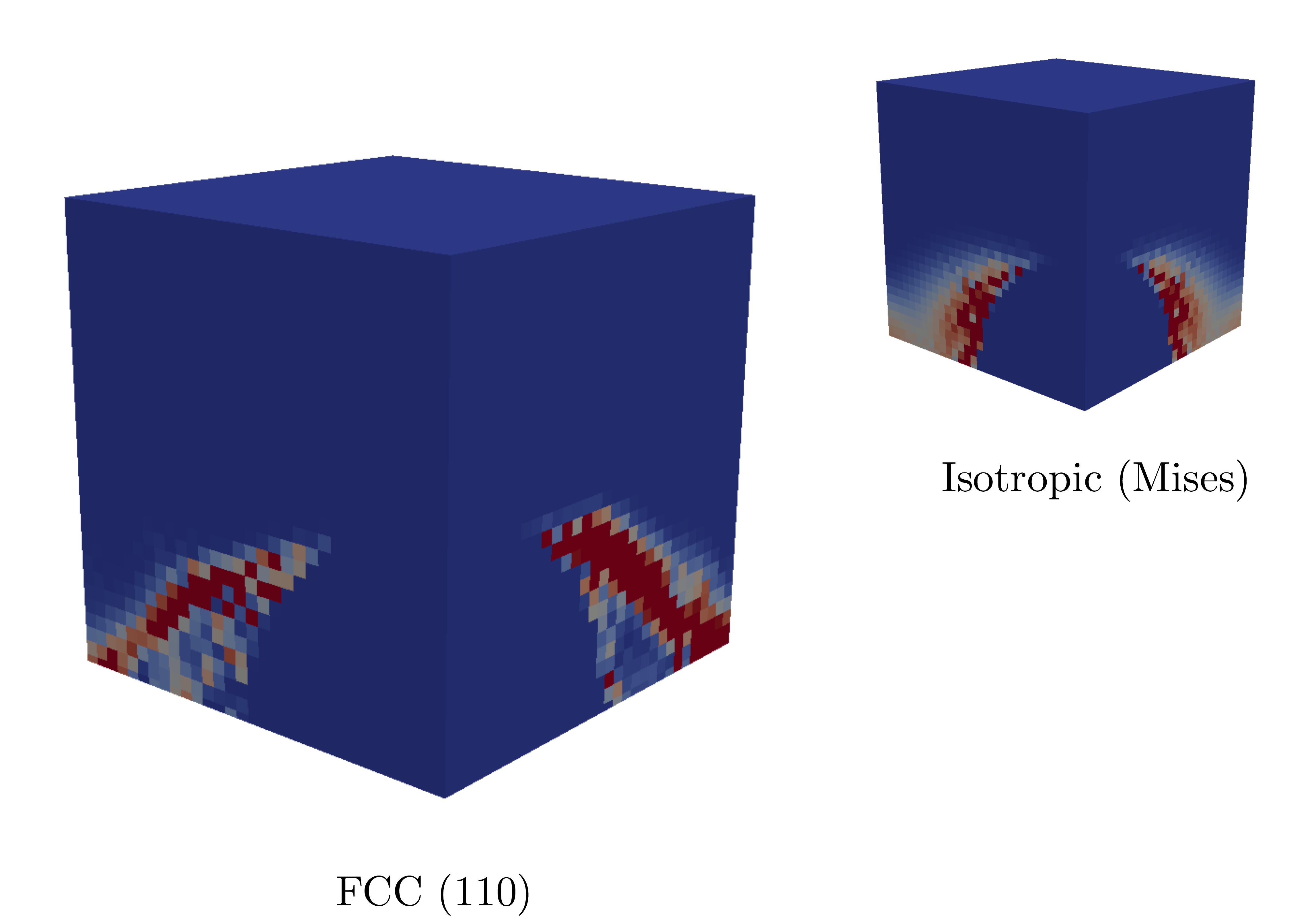}}
\subfigure[]{\includegraphics[height = 4.cm]{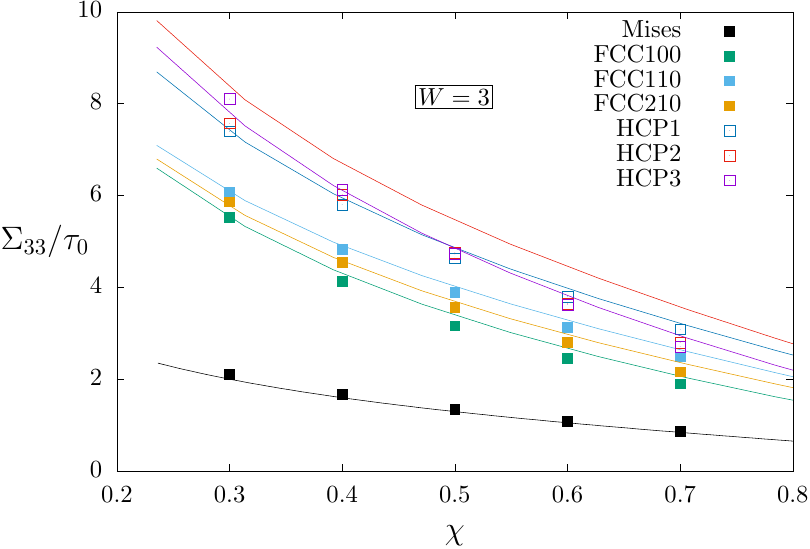}} \hspace{1.3cm}
\subfigure[]{\includegraphics[height = 4.cm]{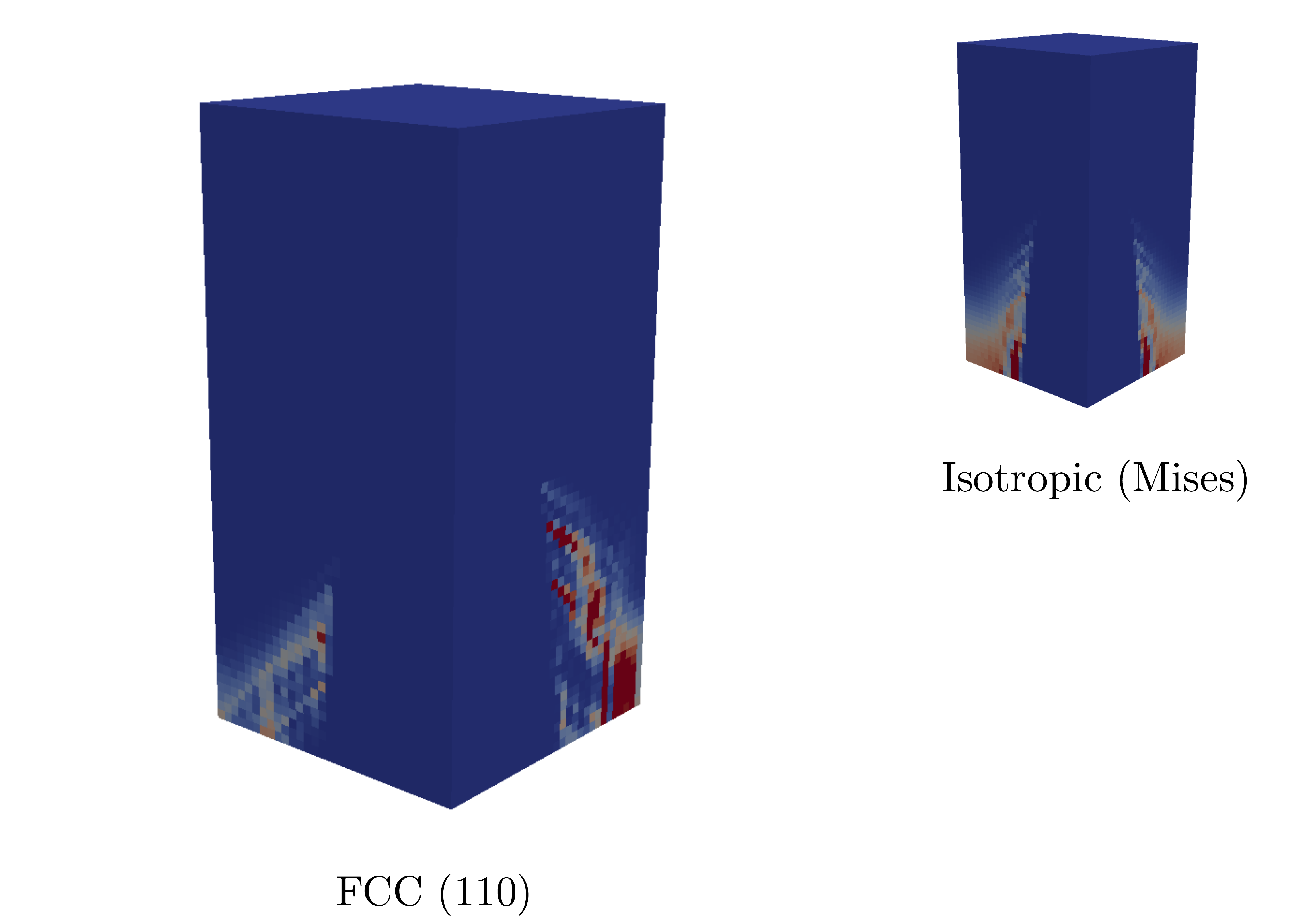}} \hspace{0.cm}
\caption{Normalised coalescence stress as a function of the dimensionless intervoid ligament size, for different constitutive equations  (von Mises, crystal plasticity (FCC, HCP) with different crystallographic orientations). Comparisons between predictions (Eq.~\ref{eqtorki},~\ref{chieff}, solid lines) and numerical results (squares), for spheroidal voids in cubic unit-cells, for $W=0.5$ (a), $W=1$ (c) and $W=3$ (e). Associated numerical equivalent strain rate fields (in arbitrary units) for FCC material oriented along $[110]$ (b,~d,~f). Insets: Numerical equivalent strain rate field for an isotropic von Mises material.}
\label{cubiaxi}
\end{figure}

As already discussed in \cite{torki}, an heuristic modification of the coalescence criterion is required to go from one configuration to the other. Based on the concept of equivalent porosity in the coalescence layer, Torki \textit{et al.} \cite{torki} proposed to consider an effective intervoid ligament:
\begin{equation}
  \chi_{eff} = \alpha \chi
  \label{chieff}
  \end{equation}
to be used in the coalescence criterion derived for cylindrical void / cylindrical unit-cell to describe the case of spheroidal void in cubic unit-cell, where $\alpha = \sqrt{\pi/6}$. In this study, the parameter $\alpha$ has been adjusted based on comparisons between numerical results and analytical predictions to recover a good agreement for a von Mises matrix material for $W=3$, leading to $\alpha = 0.85$. This parameter has then been used for all the other cases. For prolate voids $(W=3)$ (Fig.~\ref{cubiaxi}e), a good agreement is observed for FCC material between numerical results and the coalescence criterion. The criterion is able to capture quantitatively the dependence of the coalescence stress on the crystallographic orientation. The agreement is less quantitative for the case of the HCP material, where only two of the three crystallographic orientations shown are well described by the coalescence criterion. As numerical results and predictions were found to be very close for the case of cylindrical void / cylindrical unit-cell (Fig.~\ref{figresu1}e), such discrepancy is related to the effect of the unit-cell which can not be fully captured through Eq.~\ref{chieff}, especially in the case of strong anisotropy. Equivalent strain rate fields obtained in the numerical simulations are shown in Fig.~\ref{cubiaxi}f for a von Mises matrix material and FCC material (with orientations [110]). A clear effect of the anisotropy of the FCC material can be observed, but the overall field is rather close to the one obtained using von Mises plasticity, explaining the ability of the proposed criterion to describe the unit-cells results.

For spherical ($W=1$) and oblate ($W=0.5$) voids, the  coalescence criterion is found to slightly underestimate the numerical results for all situations shown on Fig.~\ref{cubiaxi}a,~c. The discrepancies are mainly due to the use of the effective intervoid ligament $\chi_{eff} = 0.85 \chi$, which has been calibrated to provide an overall agreement for isotropic von Mises material for prolate voids, but leads to underestimation for spherical and oblate voids. For the most anisotropic material (HCP), some deviations are also observed, especially regarding the influence of the crystallographic orientations on the coalescence stress, indicating a non-negligible coupling between the shape of the unit-cell and the anisotropy of the material, that can not be fully captured quantitatively through Eq.~\ref{chieff}. As for prolate voids, good agreement between numerical results and predictions are observed when the overall strain rate fields for single crystals are rather close to their isotropic counterparts (as shown on Figs.~\ref{cubiaxi}b,~d), justifying the assumption made in the derivation of the criterion. 

\subsection{Combined tension and shear loading conditions}

\begin{figure}[H]
\centering
\subfigure[]{\includegraphics[height = 3.5cm]{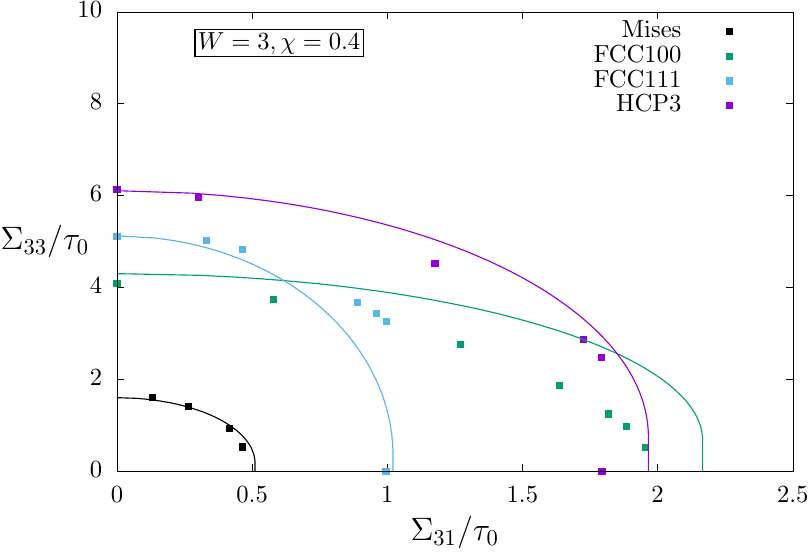}} 
\subfigure[]{\includegraphics[height = 3.5cm]{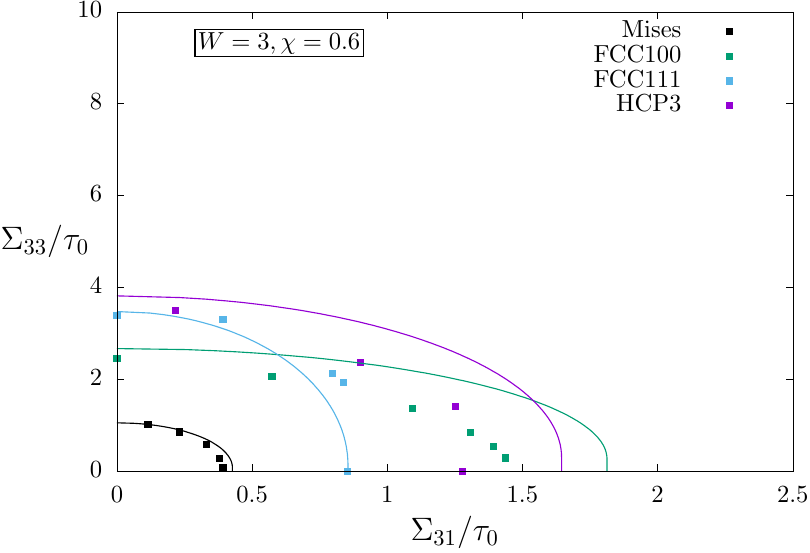}}
\subfigure[]{\includegraphics[height = 3.5cm]{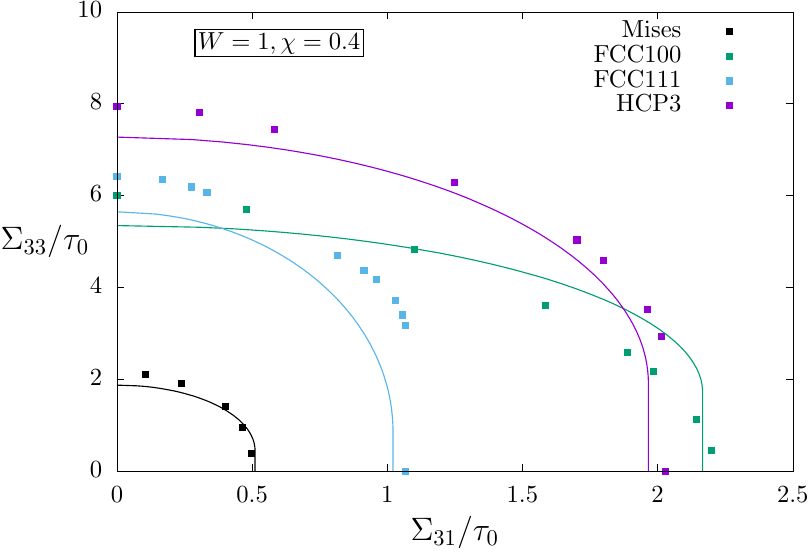}} 
\subfigure[]{\includegraphics[height = 3.5cm]{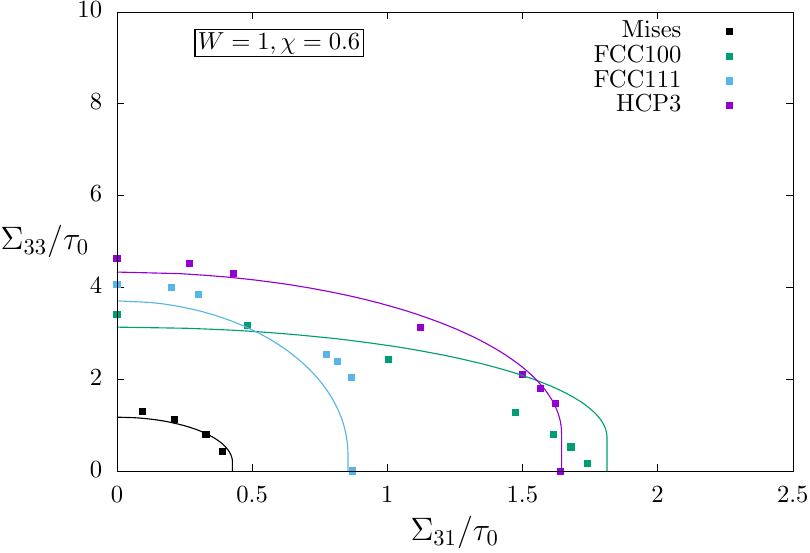}}
\subfigure[]{\includegraphics[height = 3.5cm]{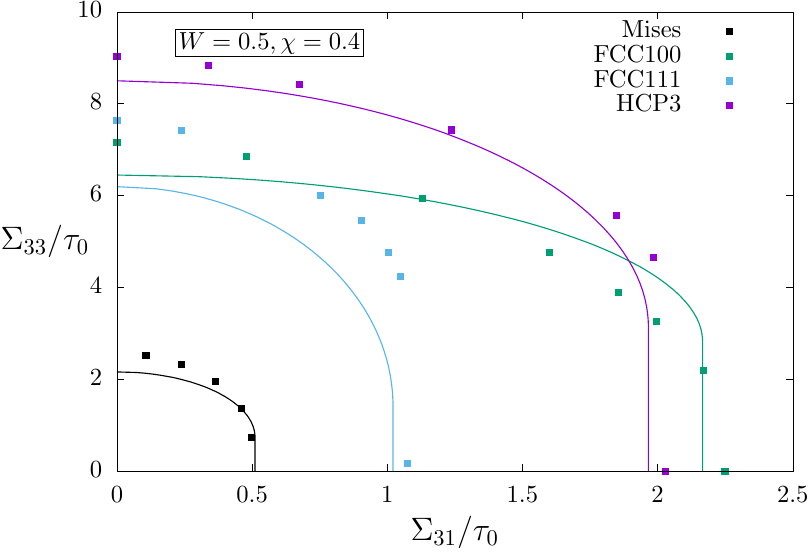}} 
\subfigure[]{\includegraphics[height = 3.5cm]{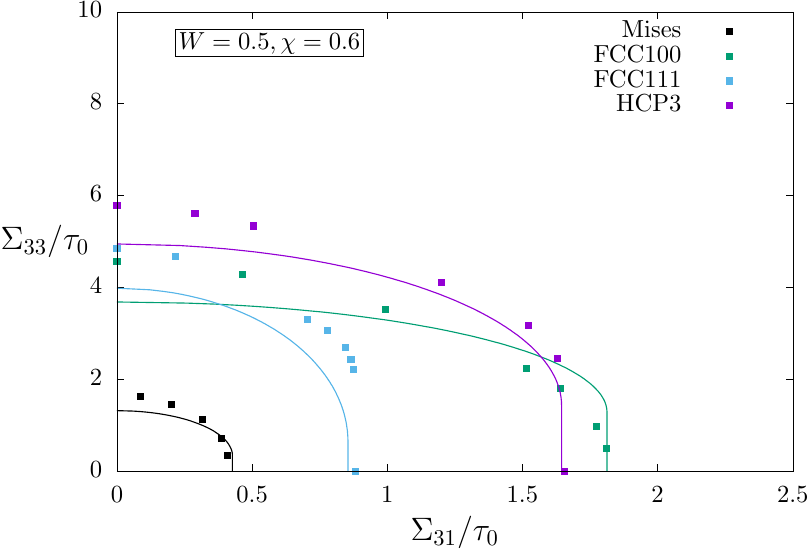}} 
\caption{Coalescence yield loci in the plane $\{\Sigma_{31},\Sigma_{33}\}$ as a function of the void aspect ratio $W \in [0.5:3]$, intervoid ligament $\chi \in [0.4;0.6]$, matrix material (von Mises, FCC, HCP) and crystallographic orientation. Solid lines correspond to Eq.~\ref{eqfullshear} and \ref{chieff}, squares to numerical results}
\label{compshear}
\end{figure}

The coalescence criterion accounting for the presence of shear with respect to the coalescence layer (Eq.~\ref{eqfullshear}) is compared in this section to the numerical results corresponding to $\alpha \neq 0$ in Eq.~\ref{loading}. Comparisons are shown on Fig.~\ref{compshear} for a von Mises matrix material and FCC100, FCC111, HCP3 single crystals, for different values of the spheroidal void aspect ratio $W$ and intervoid ligament $\chi$ for a cubic lattice of voids.

\begin{figure}[H]
\centering
\subfigure[]{\includegraphics[height = 3.8cm]{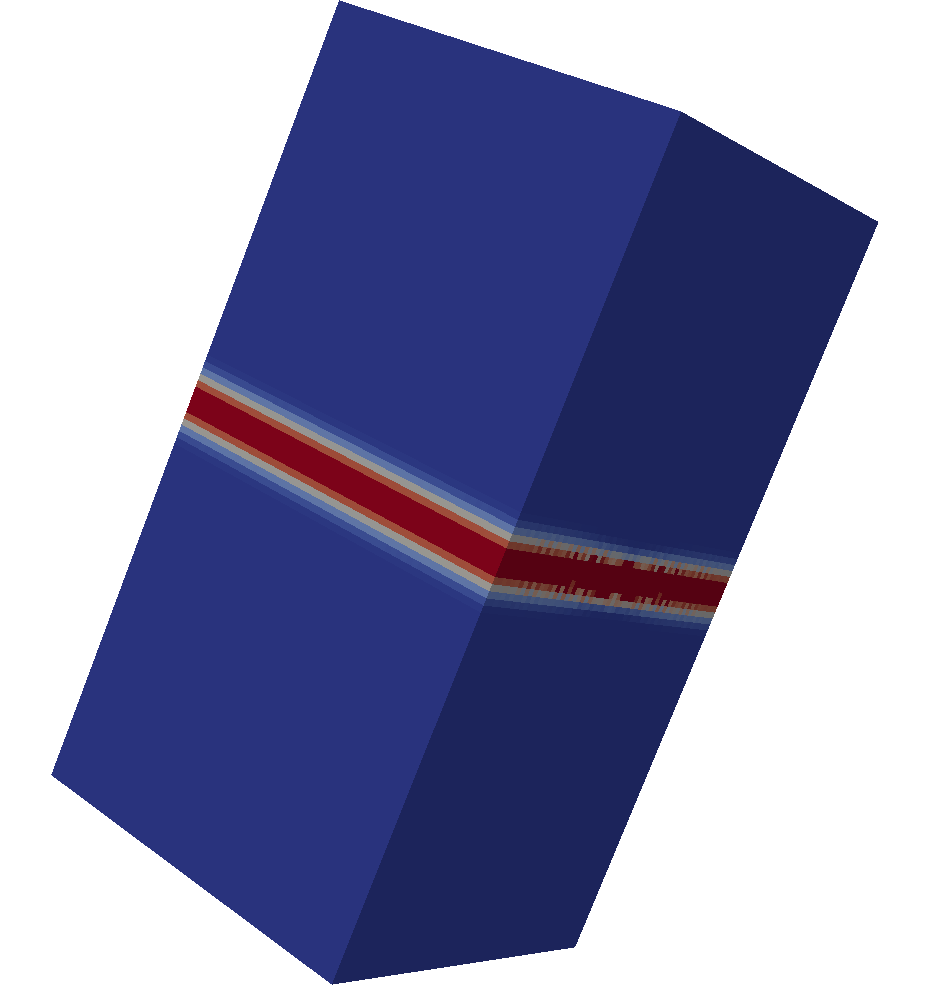}}
\hspace{0.5cm}
\subfigure[]{\includegraphics[height = 3.8cm]{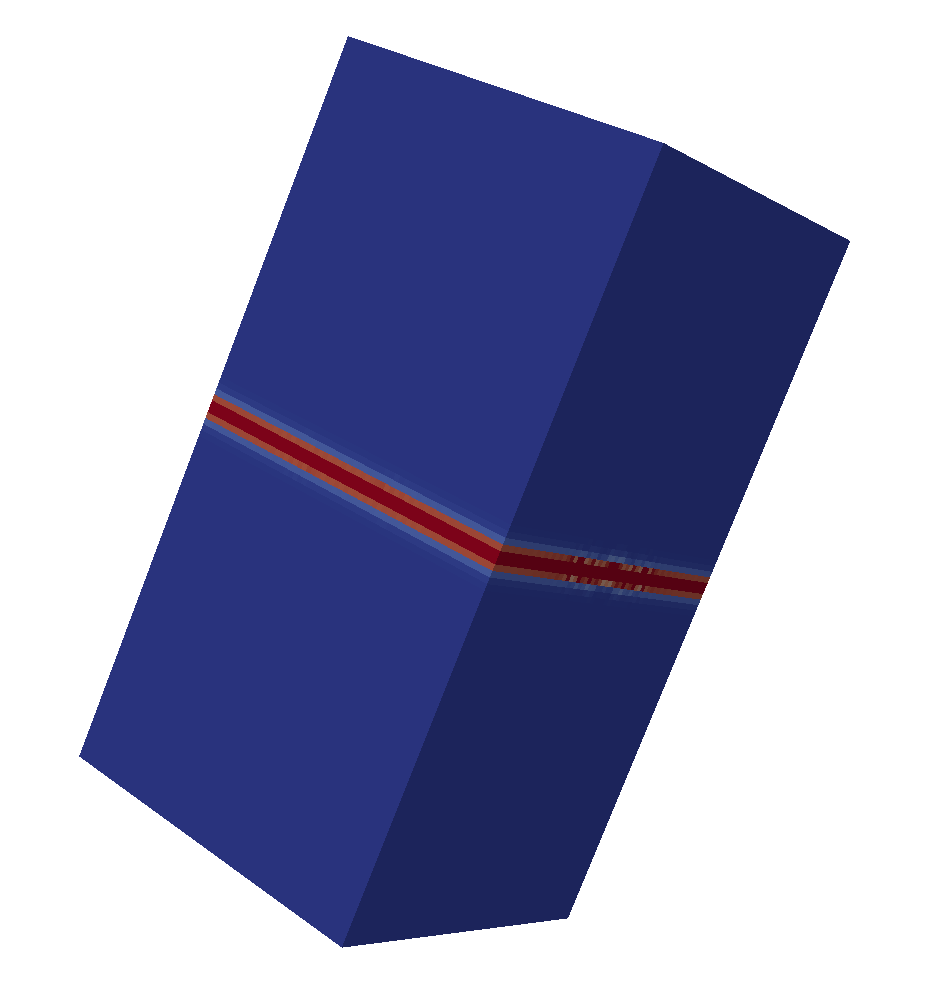}}
\hspace{0.5cm}
\subfigure[]{\includegraphics[height = 3.8cm]{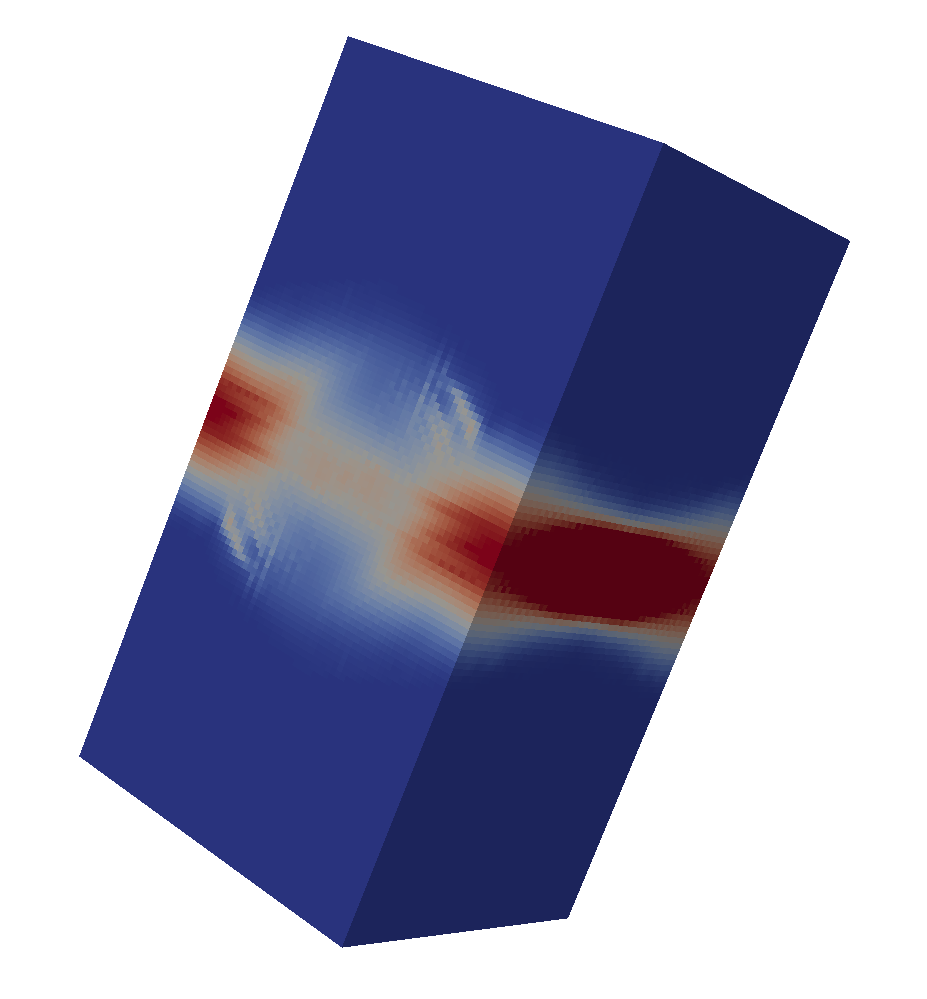}} 
\caption{Numerical equivalent strain rate fields (in arbitrary units) under pure shear loading conditions ($W=1$, $\chi=0.4$, $\Sigma_{33} \rightarrow 0$), for isotropic von Mises material (a), FCC material with orientations (b) $[111]$ and (c) $[100]$}
\label{shearfield}
\end{figure}

For the reference case of von Mises matrix material, the agreement between Eq.~\ref{eqfullshear} and numerical results is found to be good, similarly to what have been already shown in \cite{torki}. Some discrepancies appear for oblate voids for low applied shear stress, as already discussed in the previous section and attributed to the effect of the unit-cell \cite{torki2017}. Restricting to pure shear loading conditions, a strong effect of the crystallographic orientation is observed on the coalescence stress, up to a factor two for the FCC material with orientation $[100]$ and $[111]$, respectively. Such effect can be easily understood: if the coalescence layer corresponds to a slip plane of the crystal (as for the direction $[111]$), the associated shear stress is low as corresponding to a soft direction. Equivalent strain rate fields for von Mises material and FCC material with crystallographic orientations $[111]$ and $[100]$ (Figs.~\ref{shearfield}a,~b,~c, respectively) confirm this observation, showing especially the appearance of a slip band for the orientation $[111]$.


On the whole, a satisfactory agreement is observed between numerical simulations and the predictions of the coalescence criterion (Eq.~\ref{eqfullshear}). The dependences on the set of slip systems (FCC, HCP) and crystallographic orientations of coalescence yield locus are well captured. Eq.~\ref{eqfullshear} is shown to be able to reproduce quite accurately the coalescence stress under combined tension and shear loading conditions for all cases investigated, with the notable exception of prolate voids ($W=3$) and large values of the intervoid ligament ($\chi = 0.6$). Thus the discrepancies between the coalescence locus given by Eq.~\ref{eqfullshear} and numerical results result mainly on the differences described in the previous section, that comes from the effect of the void lattice (cubic \textit{vs.} hexagonal) through the choice of the unit-cell.

\begin{figure}[H]
\centering
\subfigure[]{\includegraphics[height = 5cm]{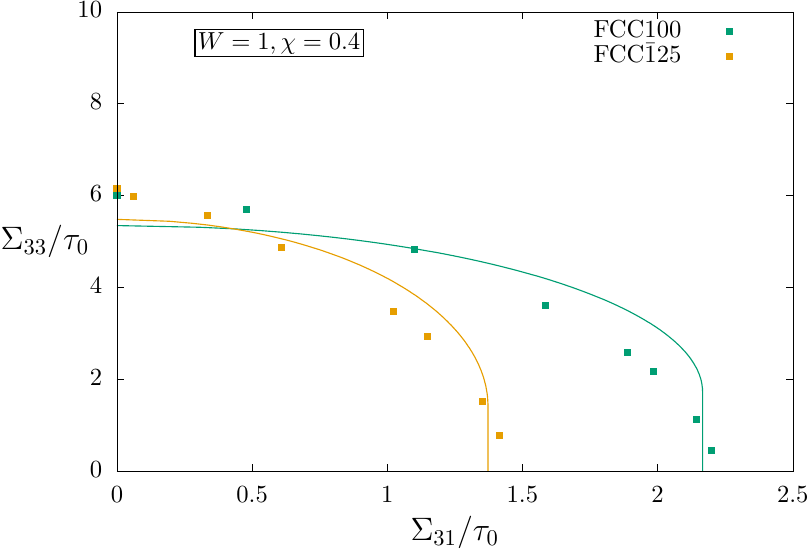}} 
\subfigure[]{\includegraphics[height = 5cm]{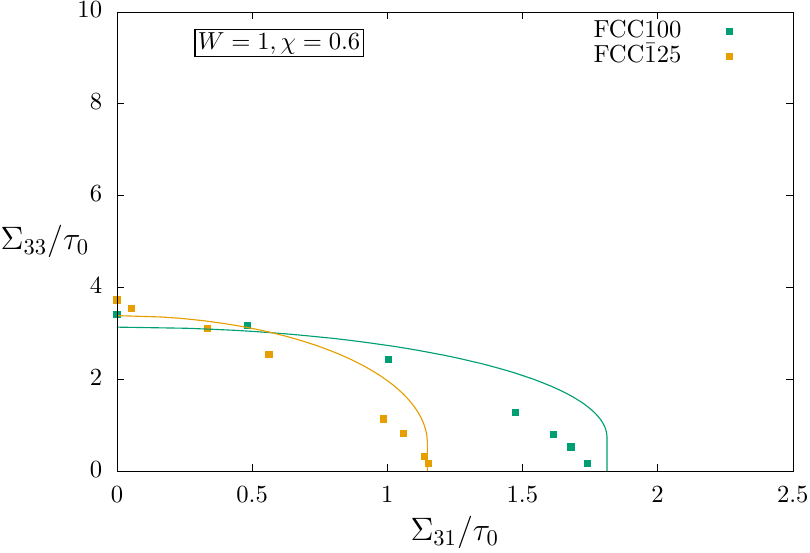}}
\caption{Coalescence yield loci in the plane $\{\Sigma_{31},\Sigma_{33}\}$ as a function of the intervoid ligament $\chi \in [0.4;0.6]$ for $W=1$, FCC matrix material with symmetric and non-symmetric crystallographic orientations. Solid lines correspond to Eq.~\ref{eqfullshear} and \ref{chieff}, squares to numerical results}
\label{compshear2}
\end{figure}

Fig.~\ref{compshear} corresponds mainly to crystal orientations exhibiting high symmetry with respect to the loading directions. In order to assess further the coalescence criterion (Eq.~\ref{eqfullshear}), numerical results for a crystal orientation with low symmetry (FCC [-125]) are compared to the predictions in Fig.~\ref{compshear2} where results from the highly symmetric orientation (FCC [100]) are also reported for comparisons. In both cases, the numerical data are well described by the proposed criterion, indicating that Eq.~\ref{eqfullshear} is also well adapted for non-symmetric crystallographic orientations. \textcolor{black}{Interestingly, the numerical coalescence locus for some crystallographic orientations shown in Figs.~\ref{compshear} or~\ref{compshear2} seems to cross the x-axis with an angle, contrary to what is observed for isotropic (Mises) material \cite{torki}. Such observation is not explained yet, and requires further assessment in a future study.}

The comparisons presented in Figs.~\ref{figresu1},~\ref{cubiaxi},~\ref{compshear} and~\ref{compshear2} show that the coalescence criterion defined by Eqs.~\ref{eqtorki} and \ref{eqfullshear} is able to capture almost quantitatively the effect of the strong crystal anisotropy on the coalescence stresses. Both dependence on the set of slip systems considered and crystallographic orientations are captured. Some discrepancies are observed for the case of a cubic lattice of spheroidal voids, that have been attributed to the difference between the geometry used for the derivation of the criterion and the cases tested, that can not be captured by the simple relation (Eq.~\ref{chieff}) used for isotropic (von Mises) material. The aim of the next section is to assess the implications of the proposed coalescence criterion for the deformation mode of porous single crystals, regarding the competition between void growth and void coalescence, as well as the coupling between void lattices and available slip systems.

\section{Discussion}

\subsection{Competition between void growth and coalescence}

\begin{figure}[H]
\centering
\subfigure[]{\includegraphics[height = 5cm]{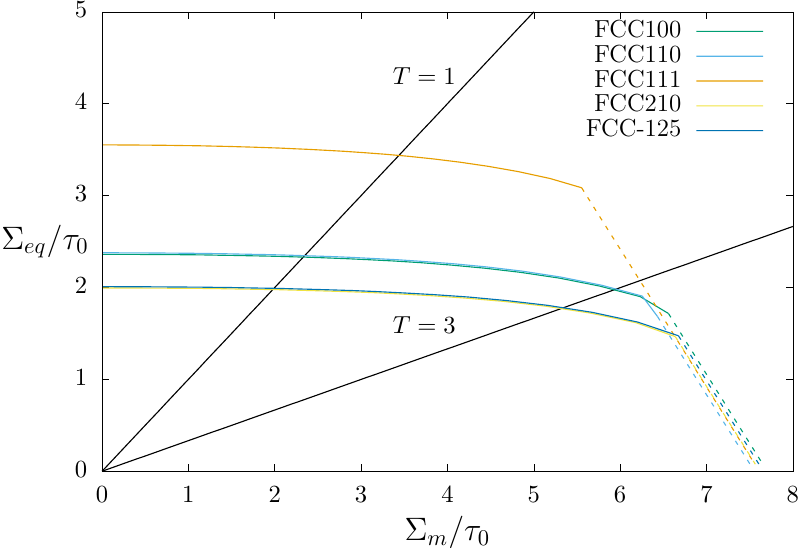}} 
\subfigure[]{\includegraphics[height = 5cm]{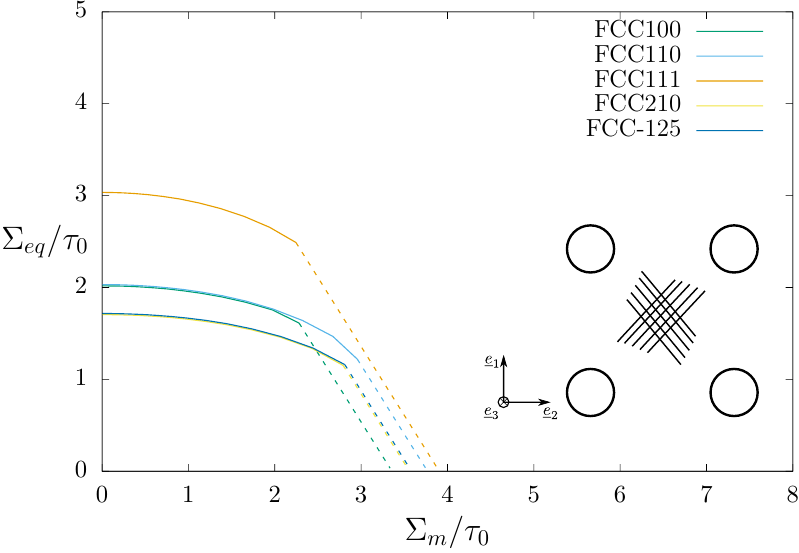}}
\caption{Yield loci of porous FCC single crystals with different crystallographic orientations, under axisymmetric stress states (Eq.~\ref{axistress}), for porosities equal to $f=0.01$ (a) and $f=0.1$ (b). Solid lines correspond to void growth regime (Eq.~\ref{eqpaux}), dashed lines to void coalescence regime (Eq.~\ref{eqtorki2}). Inset (b): simple cubic lattice with axes aligned with principal stresses axes.}
\label{locus}
\end{figure}

As shown in Section~2, porous single crystals under mechanical loading deform rather homogeneously or heterogeneously, which corresponds to the so-called void growth and void coalescence regime \cite{benzergaleblond}. A competition between these two deformation modes arises, that depends on porosity (and voids lattice), available slip systems, crystallographic orientations and loading conditions. In order to bring new insights into this two competing deformations modes, a simple cubic lattice (with axis aligned with the principal stress axis) of spherical voids under axisymmetric loading conditions is considered in the following. The macroscopic stress can be written as:
\begin{equation}
\bm{\Sigma} =
\left( \begin{array}{ccc}
\Sigma_{11} & 0 & 0 \\
0 & \eta \Sigma_{11} & 0 \\
0 & 0 & \eta \Sigma_{11} \\
\end{array}
\right)
\label{axistress}
\end{equation}
where the parameter $\eta$ characterizes the stress triaxiality $T$ - defined as the ratio of the mean stress $\Sigma_m$ to the von Mises equivalent stress $\Sigma_{eq}$ - through the relations:
\begin{equation}
  \Sigma_{eq} = \Sigma_{11}(1 - \eta) \ \ \ \ \  \ \ \ \ \ \Sigma_m = \Sigma_{11}\frac{1+2\eta}{3} \ \ \ \ \  \ \ \ \ \ T = \frac{\Sigma_{m}}{\Sigma_{eq}} = \frac{1+2\eta}{3(1-\eta)}
  \end{equation}
Different yield criteria for porous single crystals in the growth regime have been derived theoretically and assessed numerically recently \cite{xuhan,paux,mbiakop2,song1}. For the case considered here of spherical voids, all these models lead to predictions in close agreement, and the yield criterion proposed by Paux \textit{et al.} is used\footnote{A large value of the parameter $n$ should be used to recover Schmid's law for purely deviatoric loadings. A typical value of $n=100$ is taken for numerical applications.}:
\begin{equation}
  \left( \frac{ \left(\sum_k |\bm{\mu}_k:\bm{\Sigma}|^n \right)^{1/n}}{\tau_0} \right) + 2q_1 f \cosh{\left(q_2 \frac{\Sigma_m}{\tau_0}    \right)}   -1 - q_1^2 f^2 = 0
  \label{eqpaux}
\end{equation}
where $f$ is the porosity, and $q_1 = 1.59$, $q_2=0.506$ are parameters that have been obtained numerically. For these specific conditions of simple cubic lattice of voids with axis aligned with the axis of the principal stresses, coalescence is expected to occur in the plane of normal $\underline{e}_1$ corresponding to the minimal intervoid distance (highest value of $\chi$) and maximal normal stress. This corresponds to a coalescence in absence of shear stresses, and the coalescence criterion is defined through: 
\begin{equation}
  \textcolor{black}{\frac{\Sigma_{11}}{\tau_0} = M_1 t(W,\chi) \left[\frac{\chi^3 - 3\chi + 2}{3\sqrt{3} W\chi} \right]     + M_2\frac{b}{\sqrt{3}}\left[2 - \sqrt{1+3\chi^4} + \ln \frac{1 + \sqrt{1 + 3\chi^4}}{3\chi^2}    \right]}
  \label{eqtorki2}
\end{equation}
where $W=1$ for spherical voids, and where the intervoid ligament  can be related to the porosity $\chi_{eff} = 0.85 \chi = 0.85 ([6/\pi]f)^{1/3}$. For a given set of parameters - set of slip systems, crystallographic orientations, stress triaxiality - the minimum value of $\Sigma_{11}$ obtained through Eqs.~\ref{eqpaux},~\ref{eqtorki2} defines the macroscopic yield stress and the active deformation mode. Yield loci are shown on Fig.~\ref{locus} for FCC single crystals with different crystallographic orientations, for two different porosities $f=0.01$ (Fig.~\ref{locus}a) and $f=0.1$ (Fig.~\ref{locus}b).
Each of the yield loci shown on Fig.~\ref{locus} is composed of two parts: the solid lines correspond to cases where yield stress is set by the growth yield criterion (Eq.~\ref{eqpaux}), and dashed lines to yield stress set by the coalescence criterion (Eq.~\ref{eqtorki}). For low values of stress triaxiality, void growth deformation mode is activated, and a strong dependence of the yield stress on the crystallographic orientations is observed, as already discussed in \cite{xuhan,paux}. For high stress triaxiality, void coalescence can be activated, with a weaker dependence on the crystallographic orientation  as discussed in Section~4, due to the high number of slip systems for FCC single crystals. However, for a given porosity, the transition between void growth to void coalescence appear to strongly depend on crystallographic orientations, as shown on Fig.~\ref{locus} comparing $[210]$ and $[111]$ orientations. For a given stress triaxiality, the porosity for which deformation mode will switch from growth to coalescence is much lower for the orientation $[111]$ compared to $[210]$. These observations are in qualitative agreement with finite-strain unit-cells simulations performed in \cite{chaoling} showing a porosity at coalescence of about few percents for stress triaxialities of $T=1$ and $T=3$, but not in agreement regarding the dependence of the porosity at coalescence on the crystal orientations. However, direct quantitative comparisons with data provided in \cite{chaoling} are not possible, as depending on several other factors such as hardening, evolution of void aspect ratio, that are outside the scope of this study.

\subsection{Coupling between crystal orientation and void lattice}

\begin{figure}[H]
\centering
\subfigure[]{\includegraphics[height = 5cm]{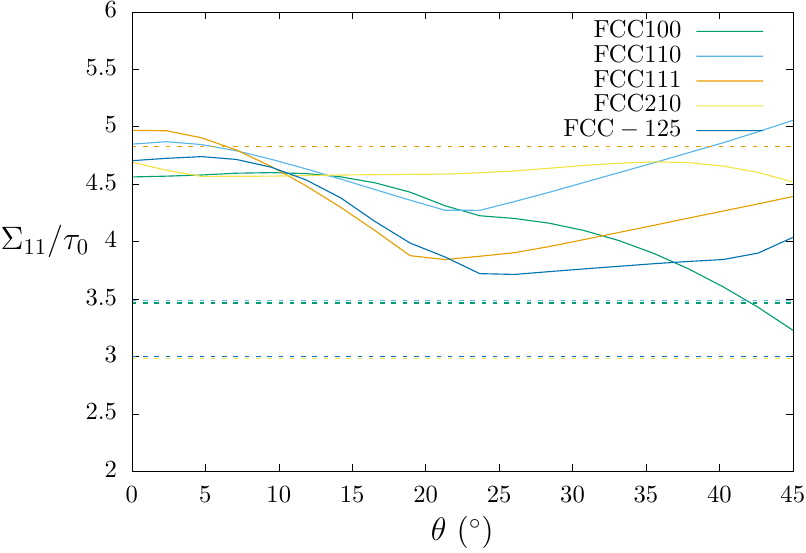}} 
\subfigure[]{\includegraphics[height = 5cm]{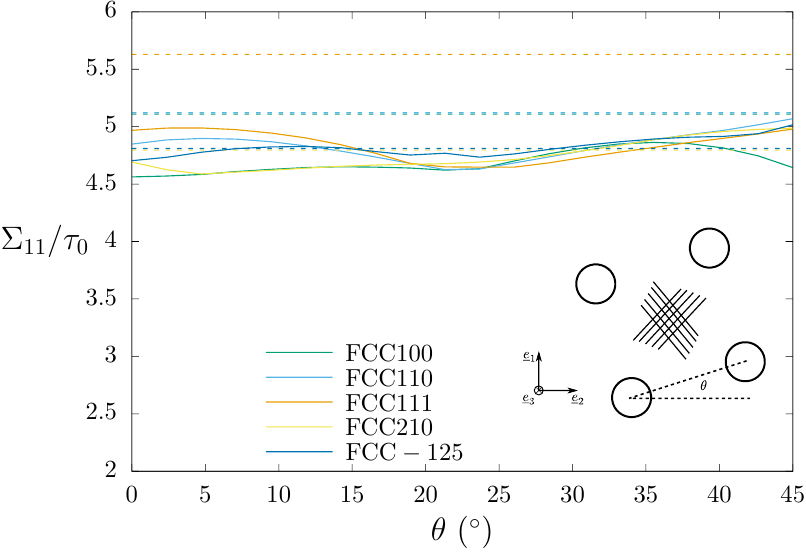}}
\caption{Yield stresses as a function of the simple cubic void lattice angle (Inset (b)) for porous FCC single crystals with different crystallographic orientations, under axisymmetric stress states (Eq.~\ref{axistress}), for porosity $f=0.05$ and stress triaxialities $T=1$ (a) and $T=3$ (b). Dashed lines correspond to void growth regime (Eq.~\ref{eqpaux}), solid lines to void coalescence regime (Eq.~\ref{eqtorki2}).}
\label{rotheta}
\end{figure}

The potential coupling between void lattice and crystal orientation is investigated in this section. A simple cubic void lattice is still considered, but with a rotation of angle $\theta$ with respect to the axis $\underline{e}_3$ (Inset Fig.\ref{rotheta}b). As in the previous section, macroscopic yield stress is computed using Eq.~\ref{eqpaux} for void growth and Eq.~\ref{eqfullshear} for void coalescence, keeping the assumption that the coalescence layer will appear along the direction where the intervoid distance is the smaller. For axisymmetric stress conditions as defined by Eq.~\ref{axistress}, normal and shear stresses on the coalescence layer, needed to use Eq.~\ref{eqfullshear}, are computed as:
\begin{equation}
  \Sigma_{nn} = \underline{n}.\bm{\Sigma}.\underline{n} \ \ \ \ \ \ \ \ \ \ \Sigma_{sh} = \sqrt{\underline{n}.(\bm{\Sigma}.\bm{\Sigma}).\underline{n} - (\underline{n}.\bm{\Sigma}.\underline{n})^2}
\end{equation}
with $\underline{n} = \{\cos{\theta}\ \ \sin{\theta}\ \ 0\}$. FCC set of slips systems are used with the five different crystallographic orientations defined in Tab.~2, for two stress triaxialities $T \in \{1,3\}$. 

Macroscopic yield stresses computed using Eq.~\ref{eqpaux} and Eq.~\ref{eqtorki} as a function of the rotation of the void lattice are shown on Fig.~\ref{rotheta}, for a porosity $f=0.05$ and for two different stress triaxialities $T=1$ (Fig.~\ref{rotheta}a) and $T=3$ (Fig.~\ref{rotheta}b). In both cases, the macroscopic stress for void growth (horizontal lines in Fig.~\ref{rotheta}) is independent of the void lattice rotation, as depending only of the porosity. On the contrary, the macroscopic stress for void coalescence depends on voids lattice, as shown especially on Fig.~\ref{rotheta}a. Depending on the rotation angle $\theta$, either void growth or void coalescence can be activated first, as for the $[111]$ orientation where void growth is expected to be the deformation mode for $\theta \leq 10^{\circ}$, while void coalescence is expected for $\theta \geq 10^{\circ}$. Such strong coupling between crystal orientation and voids lattice is more important for low stress triaxialities as shear stresses acting on the coalescence layer can reach higher values. For relatively large stress triaxiality, as in Fig.~\ref{rotheta}b where $T=3$, the effect of void lattice angle on coalescence stress is more limited.

\section{Conclusions and Perspectives}

Significant progress has been obtained in the last few years concerning the homogenization of porous single crystals, motivated by experimental observations of voids lower than the grain size in many materials or by the use of single crystals structural components. Macroscopic yield criteria have been proposed for porous single crystals with spherical voids \cite{xuhan,paux} and subsequently for spheroidal voids \cite{mbiakop2,song1}, leading to the proposition of a finite-strain homogenized elastoviscoplastic model for porous single crystals, incorporating hardening and porosity evolution \cite{chaoling}. Unit-cells simulations on periodic arrays of voids in single crystals have clearly shown that void coalescence deformation mode - characterized by localized / heterogeneous plastic flow in the intervoid ligament - appears for large value of the porosity, which is not described by the available homogenized models.

A coalescence criterion is proposed in this study for porous single crystals (where plastic flow is governed by slip only) with periodic arrays of voids. A first approach based on the quadratic approximation of the crystal plasticity yield criterion - leading to a Hill-type plastic criterion - and the use of a coalescence criterion derived for orthotropic (Hill-type) material fails to provide predictions in quantitative agreement with numerical results, although capturing qualitatively the effect of crystallographic orientations (Appendix~A). A second approach has been developed within the framework of limit-analysis and homogenization, based on the extension of a coalescence criterion derived for isotropic materials through the use of average Taylor factors parameters. Such approach leads to coalescence stress predictions in very good agreement with numerical results - obtained through numerical limit-analysis - for cylindrical voids in cylindrical unit-cells (approximating a hexagonal lattice of voids), capturing the effects of the set of slip systems considered, crystallographic orientations and void shape and size. The coalescence criterion has been further validated through comparisons to numerical results for spheroidal voids in cubic unit-cells (approximating a simple cubic lattice of voids) with or without the presence of additional shear with respect to the coalescence plane. The proposed criterion has been used to address the effect of crystallographic orientation on the transition between void growth to coalescence, as well as the strong coupling between crystal and void lattice orientations on coalescence stress, for low applied macroscopic stress triaxiality.

In order to provide a complete homogenized model for porous single crystals, several aspects of the present work should be studied in the future. First, the coalescence criterion depends on average Taylor factors which need to be computed numerically. Such computation, albeit straightforward, may be numerically inefficient when calling each time a simplex algorithm for the integrand function. However, such procedure can be made efficient numerically by pre-computing Taylor factors and using fitting functions, as already proposed and discussed in studies dealing with the macroscopic response of polycrystalline aggregates \cite{vanhoutte}. Secondly, a complete homogenized model for (periodic-)porous single crystals including void coalescence requires adding a flow rule (through normality to the criterion), evolution laws for the geometrical parameters (from volume conservation) as well as incorporating hardening (through Gurson's original proposal or in a more refined way \cite{paux2}), ultimately relying on comparisons to unit-cells simulations. \textcolor{black}{In particular, accounting for hardening is expected to be a crucial point as it was shown in \cite{yerra} that coalescence criterion depends on local yield stress.} Two more challenging numerical tasks are to find efficiently the coalescence plane for a given microstructure (void lattice and crystal orientation). Indeed, Eq.~\ref{eqfullshear} corresponds to an infinity of yield criteria associated with each possible direction for the normal and direction of the shear stress, and also to incorporate the coalescence yield criterion into a finite-strain framework, which deserve careful attention. Finally, the coalescence criterion proposed in this study relies (as well as all the other coalescence criterion proposed in the literature) on the assumption of a periodic array of voids. The relevance of these coalescence criteria for porous materials with random distribution of voids still remains to be assessed.\\  

\noindent
\textbf{Acknowledgements}\\

The author would like to thank L. G\'el\'ebart for developing the \texttt{AMITEX\_FFTP} software used in this study and for providing technical assistance, P.O. Barrioz for providing experimental data, J.M. Scherer and K. Danas for fruitful discussions, and B. Tanguy for the careful reading of the manuscript.
\newpage

\section{Appendix A: Coalescence criterion for FCC single crystals: Quadratic approximation}
\label{appendixA}
\noindent
Regularized crystal plasticity constitutive equations rely on the definition of a single yield criterion \cite{arminjon}:
\begin{equation}
  \mathcal{F} = \left[ \sum_k \left( \bm{\sigma}:\bm{\mu}^k \right)^n \right]^{1/n} - \tau_0
  \label{eqA1}
  \end{equation}
Quadratic approximation $n=2$ reduces Eq.~\ref{eqA1} to Hill-type plastic criterion:
\begin{equation}
  \mathcal{F} = \sqrt{\frac{3}{2}\bm{\sigma}:\mathbbm{p}:\bm{\sigma}} - \tau_0 \ \ \ \ \ \mathrm{with} \ \ \ \ \ \mathbbm{p} = \frac{2}{3}\sum_k \bm{\mu}^k \otimes \bm{\mu}^k
    \label{eqA2}
\end{equation}
For cubic materials, any symmetric fourth-order tensor $\mathbbm{A}$ can be written as \cite{norris}:
\begin{equation}
  \mathbbm{A} =  j \mathbbm{J} + m\mathbbm{M} + l\mathbbm{L} 
  \label{eqp}
  \end{equation}
where the fourth-order tensors $\mathbbm{L}$, $\mathbbm{M}$ and $\mathbbm{J}$ are orthogonal, idempotent, and defined such as:
\begin{equation}
\mathbbm{J} = \frac{1}{3}\textbf{I}\otimes\textbf{I}\ \ \ \ \ \ \ \ \ \ \ \ \ \ \ \mathbbm{L} = \bm{U} \otimes \bm{U} + \bm{V} \otimes \bm{V} + \bm{W} \otimes \bm{W}\ \ \ \ \ \ \ \ \ \ \ \ \ \ \ \mathbbm{M} = \frac{2}{3}(\bm{X} \otimes \bm{X} + \bm{Y} \otimes \bm{Y} + \bm{Z} \otimes \bm{Z})
\end{equation}
\begin{equation}
\bm{U} = \frac{1}{\sqrt{2}}(\underline{a}\otimes\underline{c} + \underline{c}\otimes\underline{a})\ \ \ \ \ \bm{V} = \frac{1}{\sqrt{2}}(\underline{b}\otimes\underline{c} + \underline{c}\otimes\underline{b})\ \ \ \ \ \bm{W} = \frac{1}{\sqrt{2}}(\underline{a}\otimes\underline{b} + \underline{b}\otimes\underline{a})
  \end{equation}
\begin{equation}
\bm{X} = \frac{1}{\sqrt{2}}(\underline{c}\otimes\underline{c} - \underline{a}\otimes\underline{a})\ \ \ \ \ \bm{Y} = \frac{1}{\sqrt{2}}(\underline{b}\otimes\underline{b} - \underline{c}\otimes\underline{c})\ \ \ \ \ \bm{Z} = \frac{1}{\sqrt{2}}(\underline{a}\otimes\underline{a} + \underline{b}\otimes\underline{b})
\end{equation}
\begin{equation}
  \mathbbm{L} + \mathbbm{M} + \mathbbm{J} = \mathbbm{I} \ \ \ \ \ \ \ \ \ \ \ \ \ \ \ \mathbbm{K} =  \mathbbm{L} + \mathbbm{M}
\end{equation}
where $\underline{a}$, $\underline{b}$ and $\underline{c}$ are the orthogonal unit-vectors defining the cubic material. For FCC materials with twelve slips systems $\{111\}<110>$, Eq.~\ref{eqA2} reduces to \cite{paux}:
\begin{equation}
  \mathbbm{p} = \frac{4}{3}\mathbbm{M} + \frac{4}{9}\mathbbm{L}
  \label{eqA3}
  \end{equation}
Eq.~\ref{eqA3} coincides with the unregularized crystal plasticity criterion only for specific orientations. A closer agreement, in an average sense, can be obtained with \cite{paux}:
\begin{equation}
  \mathbbm{p} = \frac{1}{6}\mathbbm{M} + \frac{2}{27}\mathbbm{L}
  \label{eqA33}
  \end{equation}
The associated equivalent strain rate can be written as:
\begin{equation}
  d_{eq} = \sqrt{\frac{2}{3}\textbf{d}:\hat{\mathbbm{p}}:\textbf{d}}\ \ \ \ \ \mathrm{with} \ \ \ \ \ \hat{\mathbbm{p}} = 6\mathbbm{M} + \frac{27}{2}\mathbbm{L} 
\end{equation}
where the identity $\mathbbm{p}:\hat{\mathbbm{p}} = \hat{\mathbbm{p}}:\mathbbm{p}  = \mathbbm{K}$ has been used \cite{benzergaanisotrope}. Another definition of the equivalent strain rate is classically used:
\begin{equation}
   d_{eq} = \sqrt{\frac{2}{3}\textbf{d}:\hat{\mathbbm{h}}:\textbf{d}}\ \ \ \ \ \mathrm{with} \ \ \ \ \ \hat{\mathbbm{p}} = \hat{\mathbbm{K}}:\hat{\mathbbm{h}}:\hat{\mathbbm{K}}
  \end{equation}
The components of the Voigt-Mandel representation of the fourth-order tensor $\hat{\mathbbm{h}}$ are used to compute scalar anisotropy factors \cite{keralavarma}:
\begin{equation}
\left\{
\begin{aligned}
\hat{h}_q &= \frac{\hat{h}_{11} + \hat{h}_{22} + 4\hat{h}_{33} - 4\hat{h}_{23} - 4\hat{h}_{31} + 2\hat{h}_{12}}{6} \\
\hat{h}_{t} &= \frac{\hat{h}_{11}+\hat{h}_{22}+2\hat{h}_{66} - 2\hat{h}_{12}}{4}\\
\hat{h}_{a} &= \frac{\hat{h}_{44}+\hat{h}_{55}}{2}
\end{aligned}
\right.
\label{hqhaht}
\end{equation}
These scalar anisotropy factors are computed for the FCC crystallographic orientations used in this study in Tab.~\ref{scalar}. A coalescence criterion has been proposed in \cite{keralavarma} for materials obeying Hill plastic criterion under axisymmetric loading conditions:
\begin{equation}
\left(   \frac{\Sigma_{33}}{\sigma_0}\right) \approx \sqrt{\frac{6}{5} \hat{h}_q} \left[b \ln\frac{1}{\chi^2} + \sqrt{b^2+1} - \sqrt{b^2 + \chi^4} + b \ln\left(\frac{b + \sqrt{b^2 + \chi^4}}{b + \sqrt{b^2 + 1}}    \right)                \right]
\label{eqkeralavarma2}
\end{equation}
with $\displaystyle{b^2 = \frac{\hat{h}_t}{3\hat{h}_q} + \alpha \frac{\hat{h}_a}{3\hat{h}_q} \frac{5}{8W^2 \chi^2}}$, $\alpha = [1 + \chi^2 - 5\chi^4 + 3\chi^6]/12$. An empirical modification has also been proposed in \cite{keralavarma} to account for penny-shaped voids: $W \chi \leq 2W_0 \Rightarrow W \chi = W_0 + [W\chi]^2/[4W_0]$, with $W_0 = 0.12$.\\

\begin{figure}[H]
\centering
\subfigure[]{\includegraphics[height = 5cm]{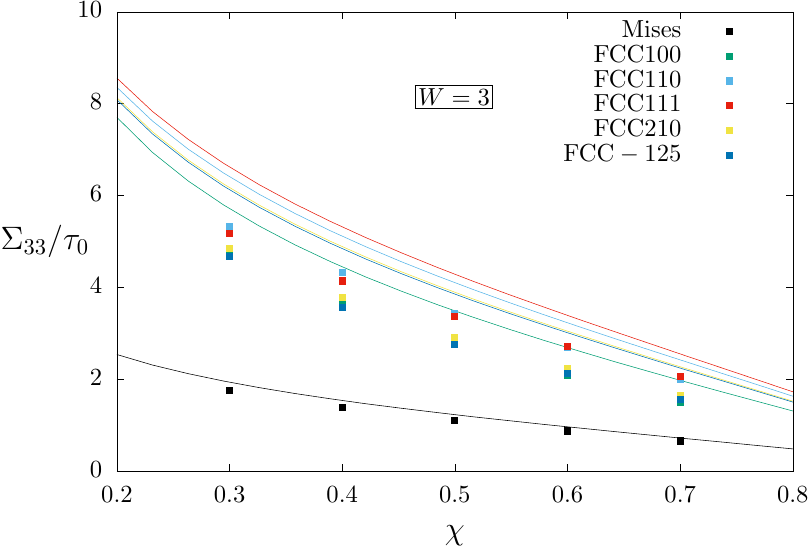}} 
\subfigure[]{\includegraphics[height = 5cm]{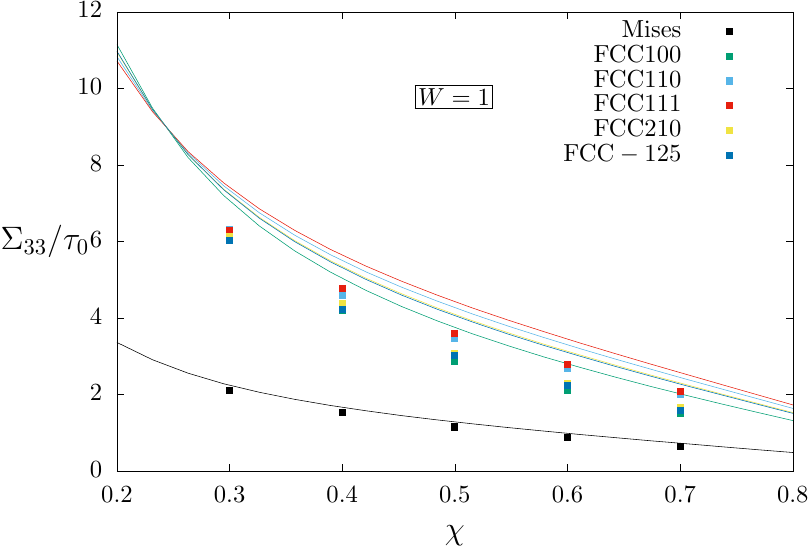}}

\caption{Normalized coalescence stress as a function of the dimensionless intervoid ligament, for different constitutive equations  (von Mises, crystal plasticity (FCC with different crystallographic orientations). Comparisons between predictions (Eq.~\ref{eqkeralavarma2}, solid lines) and numerical results (squares), for cylindrical voids in cylindrical unit-cells, for (a) $W=1$ and (b) $W=3$.}
\label{eqhill}
\end{figure}

\begin{table}

\begin{tabular}{c||c|c|c||c|c|c}
    Crystal system & \multicolumn{3}{c}{Orientation} & \multicolumn{3}{c}{Anisotropy} \\
    &  $e_3$ & $e_2$ & $e_1$ & $\hat{h}_q$ & $\hat{h}_t$ & $\hat{h}_a$\\
       \hline
       \hline
       FCC & [100]  & [001]   & [010]  & 6 & 9.75 & 13.5 \\
       FCC & [110]  & [001]   & [-110]  & 11.63 & 10.69 & 9.75 \\
       FCC & [111]  & [0-11]   & [-211] & 13.5 & 11 &  8.5\\
       FCC & [210]  & [001]   & [-120]  & 9.6  & 10.35 & 11.1\\
       FCC & [-125]  & [210]  & [1-21]  & 9.23 & 10.29  & 11.35 \\        \hline
\end{tabular}
\caption{Scalar anisotropy factors for FCC material as a function of the crystallographic orientation.}
\label{scalar}
\end{table}
Analytical predictions from Eq.~\ref{eqkeralavarma2} are compared in Fig.~\ref{eqhill} to numerical results for cylindrical voids in cylindrical unit-cells (see Section~4). A good agreement is observed for von Mises material (for which $\hat{h}_a=\hat{h}_q=\hat{h}_t=1$), as shown in \cite{keralavarma}, but Eq.~\ref{eqkeralavarma2} fails to predict quantitatively coalescence stresses for single crystals, although qualitatively capturing the effect of crystallographic orientation. \textcolor{black}{As it was shown in \cite{keralavarma} that Eq.~\ref{eqkeralavarma2} leads to rather accurate predictions for several (Hill) plastic anisotropies, discrepancies observed in Fig.~\ref{eqhill} are suspected to come from the regularization of crystal plasticity.}

\bibliographystyle{elsarticle-num.bst}
\bibliography{spebib2}

\end{document}